\documentclass[12pt]{JHEP3}

\usepackage{epsfig}
\epsfclipon
\usepackage{multicol}
\usepackage{epsfig,bm}
\usepackage{amssymb,amsmath}

\newcommand{\roughly}[1]{\mathrel{\raise.3ex\hbox{$#1$\kern-0.85em
\lower1ex\hbox{$\sim$}}}}

\newcommand{\sixteenb}{\hbox{\bf 16}}

\newcommand{\sixtyfourb}{\hbox{\bf 64}}
\newcommand{\eightyb}{\hbox{\bf 80}}

\def\be{\begin{equation}}
\def\beq\begin{equation}
\def\ee{\end{equation}}
\def\bea{\begin{eqnarray}}
\def\eea{\end{eqnarray}}
\def\hf{\frac12}
\def\pref#1{(\ref{#1})}

\def\beq{\begin{equation}}
\def\eeq{\end{equation}}
\def\beqa{\begin{eqnarray}}
\def\eeqa{\end{eqnarray}}

\def\d{\tilde{d}}

\def\Y{Y^m_l}

\newcommand{\bmat}{\left(\begin{array}}
\newcommand{\emat}{\end{array}\right)}

\def\nisubsubsection#1{\medskip\noindent {\bf #1} \smallskip\noindent}

\def\yzero{\smash{\hbox{$y\kern-4pt\raise1pt\hbox{${}^\circ$}$}}}

\def\-{\hphantom{-}}

\def\s2{\frac{1}{2}}

\def\tr{{\rm tr \,}}
\def\Tr{{\rm Tr \,}}

\def\IF{\relax{\rm I\kern-.18em F}}
\def\II{\relax{\rm I\kern-.18em I}}
\def\IP{\relax{\rm I\kern-.18em P}}
\def\IC{\relax{\rm I\kern-.48em C}}
\def\IR{\relax{\rm I\kern-.18em R}}
\def\IK{\relax{\rm I\kern-.20em K}}
\def\IM{\relax{\rm I\kern-.25em M}}

\def\cA{{\cal A}}

\def\cN{{\cal N}}

\def\Dsl{\,\raise.15ex\hbox{/}\mkern-13.5mu D} %this one can be subscripted
\def \one{\relax{\rm 1\kern-.26em I}}

\def\cR{{\cal R}}

\def\nott#1{\setbox0=\hbox{$#1$}                % set a box for #1
   \dimen0=\wd0                                 % and get its size
   \setbox1=\hbox{/} \dimen1=\wd1               % get size of /
   \ifdim\dimen0>\dimen1                        % #1 is bigger
      \rlap{\hbox to \dimen0{\hfil/\hfil}}      % so center / in box
      #1                                        % and print #1
   \else                                        % / is bigger
      \rlap{\hbox to \dimen1{\hfil$#1$\hfil}}   % so center #1
      /                                         % and print /
   \fi}                                         %

\def\L{\mathcal{L}}
\def\g{\sqrt{-g}}
\def\Y{Y_{MN}}
\def\y2{Y_{MN} Y^{MN}}
\def\Riem2{R_{MNPQ} R^{MNPQ}}
\def\altRiem2{R_{MNPQ}^{\, 2}}
\def\Ricci2{R_{MN} R^{MN}}
\def\F{F^a_{MN}}
\def\f2{F^{a}_{MN} F^{MN}_a}
\def\altf2{F_{MN}^{\, 2}}
\def\hf{\frac{1}{2}}
\def\e{\frac{1}{e}\,}
\def\nn{\nonumber}

\title{Ultraviolet Sensitivity in Higher Dimensions}

\author{D. Hoover${}^1$ and C.P. Burgess${}^{1,2,3}$\\

${}^1$ Physics Department, McGill University, 3600 University
Street,\\ \qquad Montr{\'e}al, Qu{\'e}bec, Canada, H3A 2T8. \\
${}^2$ Department of Physics and Astronomy, McMaster University,\\
\qquad 1280 Main Street West, Hamilton, Ontario, Canada, L8S 4M1.\\
${}^3$ Perimeter Institute, 31 Caroline Street North,\\ \qquad
Waterloo, Ontario, Canada.

}

\date{}
\abstract{We calculate the first three Gilkey-DeWitt (heat-kernel)
coefficients, $a_0$, $a_1$ and $a_2$, for massive particles having
the spins of most physical interest in $n$ dimensions, including
the contributions of the ghosts and the fields associated with the
appropriate generalized Higgs mechanism. By assembling these into
supermultiplets we compute the same coefficients for general
supergravity theories, and show that they vanish for many
examples. One of the steps of the calculation involves computing
these coefficients for massless particles, and our expressions in
this case agree with -- and extend to more general background
spacetimes -- earlier calculations, where these exist. Our results
give that part of the low-energy effective action which depends
most sensitively on the mass of heavy fields once these are
integrated out. These results are used in hep-th/0504004 to
compute the sensitivity to large masses of the Casimir energy in
Ricci-flat 4D compactifications of 6D supergravity.}


\begin{document}

%\newpage

%===================================================================================

\section{Introduction}

It is an experimental fact that Nature comes to us with many
scales, and that we do not need to understand them all at once in
order to understand the physics of any particular scale. Indeed,
progress on atomic physics did not have to await a complete theory
of nuclei, quarks or any hitherto-undiscovered more microscopic
constituents, and this fact arguably is fundamental to the very
possibility of making progress in science. This elementary
physical fact is reflected in the mathematics used to describe the
physical world
--- quantum field theory --- through the calculus of
renormalization and effective field theories.

Although low-energy physics is largely insensitive to higher
energy scales, it is not completely so. After all, the electronic
properties of atoms {\it do} depend on the total charge and mass
of the underlying nucleus. The calculus of renormalization, which
has become very well-developed over the last few decades, allows
the very efficient calculation of the comparatively few ways in
which short-distance high-energy physics can affect the physics of
longer wavelengths and lower energies. It does so by identifying
the low-energy effective field theory which captures the effects
of integrating out high-energy modes, and in particular finding
which effective interactions are `Ultraviolet (UV) Sensitive'
inasmuch as they are proportional to positive powers of the large
energy scale, $m$, of the particles which have been integrated out
\cite{ETbooks,EffTheories}. It is this existence of UV sensitive terms in
the low-energy effective action which underlies the `naturalness'
problems of otherwise-successful theories like the Standard Model,
including the problems of the Electroweak Hierarchy or of the
Cosmological Constant.

Although the techniques for computing UV sensitive interactions is
very highly developed for four-dimensional theories, less has been
done to compute such terms in higher-dimensional models. The
absence of such higher-dimensional results is becoming more of a
hindrance given that extra-dimensional ideas are playing an
increasingly prominent role in our understanding of the various
hierarchy problems \cite{ADD,RS,SLED}. Fortunately, well-developed
heat-kernel techniques exist for computing UV sensitivity for
reasonably general geometries \cite{Gilkey}, and it is the purpose
of this paper to use these techniques to provide a systematic
calculation of the leading sensitivity to heavy masses (within the
one-loop approximation) in higher-dimensional theories.

In order to do so we compute the most UV sensitive contributions
which are obtained when massive particles are integrated out at
one loop. We calculate the leading heat-kernel coefficients for
$n$ spacetime dimensions and for a broad class of particle spins,
including most particle types which arise within the
higher-dimensional supergravities which are of the most modern
interest. Similar heat-kernel calculations have been performed in
the past for massless particles
\cite{PreviousHeatKernel,Duffcc,Tseytlin}, and more recently for
certain massive fields in 4D \cite{HKmassive}. The results
presented herein extend these earlier calculations in several
ways. Our main extension is to provide the $n$-dimensional results
for massive fields rather than massless ones, including
calculating the contributions of the various ghosts and would-be
Goldstone particles which participate in the generalized
higher-spin mass-acquisition (Higgs) mechanisms. As an
intermediate step we also compute the leading heat-kernel
coefficients for massless particles, extending previous general
results to include a nonzero cosmological constant in
$n$-dimensions. An application of these results to the study of UV
sensitivity in Ricci-flat 4D compactifications of 6D supergravity
may be found in ref.~\cite{Doug}.

Our calculations are presented as follows. The next section, \S2,
summarizes the general heat-kernel formulae and evaluates them for
the various massless fields which arise within higher-dimensional
supergravities. These calculations are performed in a covariant
gauge, for which the gauge-fixing and ghost contributions are
explicitly displayed. For higher-spin fields the generalizations
to nonzero masses are computed by coupling the massless fields to
the appropriate would-be-Goldstone fields, whose eating makes up
the generalized Higgs mechanism for the fields of interest. \S3
then applies the general results of \S2 to the field content of
specific supergravities. As a check, and in order to compare with
previous results, the contributions of the massless fields of 10-
and 11-dimensional supergravities are computed and shown to sum to
zero, in agreement with earlier calculations. The contributions of
massive fields and supermultiplets in 4, 6 and 10 dimensions are
also tabulated in this section.

\section{General One-Loop Results}

This section collects the results for the most
ultraviolet-sensitive parts of the one-loop action obtained by
integrating out massless and massive particles having spins up to
and including spin two. For the present purposes we take the
one-loop approximation to represent the field theory which is
obtained by linearizing the various field equations about a
particular background configuration. That is, denoting the set of
(real) quantum fields generically by $\Phi^i$, with background
value $\varphi^i$, we write $\Phi^i = \varphi^i + \phi^i$ and
expand the classical action to quadratic order in $\phi^i$:
\beq
    S \approx - \, \frac12 \int d^nx \; \phi^i \,
    \Delta_{ij}(\varphi) \phi^j \,.
\eeq
Here $n$ denotes the dimension of spacetime, and in practice we
consider nonzero backgrounds only for scalar, gauge and
gravitational fields. We do, however, allow {\it fluctuations}
about these backgrounds for all of the most commonly encountered
fields in higher-dimensional supergravity theories.

\subsection{The Gilkey-DeWitt Coefficients}
\label{GilkeyDeWitt}

The full one-loop quantum correction to the effective action,
$\Sigma$, in the presence of various background fields can be
explicitly calculated provided one can evaluate the functional
determinant of the relevant differential operator in the presence
of those backgrounds. For a basis of real fields, $\phi^j$, whose
linearized equation of motion is ${\Delta^i}_j \, \phi^j$ the one
loop contribution to the effective action is
\beq
    \label{eqn: gamma1}
    i\Sigma = - (-)^{F} \frac{1}{2}\, \Tr \log
    \Delta \,,
\eeq
where $F$ denotes the fermion number of these fields (which is odd
for fermions and even for bosons). Unfortunately the evaluation of
the right-hand side of this expression is in general quite
difficult, and explicit results are typically known only for
background fields which are sufficiently simple.

Calculations are easier if one is only interested in those parts
of $\Sigma$ which are the most sensitive to very short-distance
physics. In this case very general results can be obtained by
using the Gilkey-DeWitt heat-kernel methods. For instance, the
parts of $\Sigma$ which depend the most strongly on the mass
matrix, $m$, (in the limit that the eigenvalues of $m$ are large)
can be written as
\beq \label{eqn: heatkernel}
    \Sigma_{UV} = \frac{1}{2}(-)^{F} \left(
    \frac{1}{4 \pi} \right)^{n/2} \int d^n x \sqrt{-g}
    \sum_{k=0}^{[n/2]} \Gamma ( k - n/2) \, \tr[ m^{n-2k}\, a_k ]
\eeq
where $g$ is the determinant of the metric, $\Gamma(z)$ is Euler's
gamma function and $n$ is the number of spacetime dimensions. The
$a_k$ are local quantities constructed from $k$ powers of the
background curvature tensor, as well as of the other background
fields. We stop the sum at $k = [n/2]$ --- where $[n/2]$ denotes
the largest integer which is $\le n/2$ --- since our interest is
only in those terms which do not involve negative powers of $m$.
Although all $[n/2]$ coefficients $a_k$ are required to completely
describe the UV properties of an $n$-dimensional theory, for
practical reasons we calculate here only the first three (the
number of terms in each $a_k$ increases exponentially with $n$,
c.f. eqs.~\pref{eqn: gilkey} and ~\pref{a3}). Potential
ultra-violet divergences in this expression are regulated by
taking $n$ to approach continuously the integer value of interest.

Very general explicit expressions for the first few $a_k$ are
known in some circumstances. Consider, for example, $N$ real
fields, $\phi^i$, whose field equation when linearized about the
background configuration is
\beq
    \Delta^i_{j} \phi^j = (-\Box + m^2 + X)^i_{j} \phi^j = 0 \,,
\eeq
where $\Box = g^{MN}D_M D_N$ is constructed from
background-covariant derivatives, $D_M$, and the quantity $X^i_j$
is a local background-field dependent quantity. Using the heat
kernel expansion, it is possible to show that the first few $a_k$,
are given by:\footnote{Our metric is `mostly plus' and we adopt
Weinberg's curvature conventions \cite{GandC} (which differ from
those of Misner Thorne and Wheeler \cite{MTW} only in the overall
sign of the curvature tensors).}
\begin{eqnarray}
  \label{eqn: gilkey}
  a_0 &=& I \nn \\
  a_1 &=& -\frac{1}{6}(RI+6X)  \nn \\
  a_2 &=& \frac{1}{360} \left( 2 \Riem2 - 2 \Ricci2 + 5 R^2 -12\, \Box R
  \right) I \nonumber \\  &&+ \frac{1}{6} R X + \frac{1}{2} X^2 -
  \frac{1}{6} \Box X + \frac{1}{12} \y2
\end{eqnarray}
and
\beqa
  \label{a3}
  a_3 &=& \frac{1}{7!} \left( - 18 \Box^2 R + 17 D_M R D^M R
  -2 D_L R_{MN} D^L R^{MN}
  -4 D_L R_{MN} D^N R^{ML} \phantom{\frac12} \right. \nonumber\\
  && + 9 D_K R_{MNLP} D^K
    R^{MNLP} +28 R \Box R - 8 R_{MN} \Box R^{MN}
    +24 {R^M}_{N} D^L D^N R_{ML} \nonumber\\
    &&+ 12 R_{MNLP} \Box R^{MNLP}
    - \frac{35}{9} \, R^3 + \frac{14}{3} \, R \,\Ricci2
    - \frac{14}{3} \, R \, \Riem2 \nonumber \\
    && + \frac{208}{9} \, {R^M}_N \, R_{ML} \, R^{NL}
    - \frac{64}{3} \, R^{MN} \, R^{KL} \, R_{MKNL}
    + \frac{16}{3} \, {R^M}_N \, R_{MKLP} \, R^{NKLP} \nonumber\\
    && \left. - \frac{44}{9} \, {R^{AB}}_{MN} \, R_{ABKL}
    \, R^{MNKL} - \frac{80}{9} \, R^{A \ M}_{\ B \ \, N} \,
    R_{AKMP} \, R^{BKNP} \right) I \nonumber \\
    &&+ \frac{1}{360} \left( 8 D_M Y_{NK} \, D^M Y^{NK}
    +2 D^M Y_{NM} \, D_K Y^{NK} + 12 Y^{MN} \Box Y_{MN}
    \phantom{\frac12} \right. \nn \\
    && - 12 {Y^M}_N \, {Y^N}_K \, {Y^K}_M - 6 R^{MNKL} \, Y_{MN}
    \, Y_{KL} +4 {R^M}_N \, Y_{MK} \, Y^{NK} \nonumber \\
    && - 5 R \, Y^{MN} \, Y_{MN} - 6 \Box^2 X + 60 X \Box X
    +30 D_M X \, D^M X - 60 X^3
    \nonumber \\
    && - 30 X \, Y^{MN} \, Y_{MN} + 10 R \, \Box X + 4 R^{MN}
    \, D_M D_N X +12 D^M R \, D_M X
    -30 X^2 \, R \nonumber \\
    && \left. \phantom{\frac12}
    + 12 X \, \Box R - 5 X \, R^2 + 2 X \, \Ricci2
    -2 X \, \Riem2 \right) \,,
\eeqa
where $I$ is the $N \times N$ identity matrix for the space of
fields of interest, and $\Y$ is the matrix-valued quantity defined
by the expression ${{Y_{MN}}^i}_j \, \phi^j = [D_M, D_N] \phi^i$.
$\Y$ may be expressed explicitly in terms of the Riemann tensor
and any background gauge fields, $A_M^a$, as:
\beq \label{eqn: Y}
    \Y = -i F^a_{MN} t_a - \frac{i}{2}
    R^{\hspace{14pt}AB}_{MN} J_{AB},
\eeq
where $t_a$ and $J_{AB}$ are the field-appropriate matrices which
generate gauge and Lorentz transformations, and $F^a_{MN}$ is the
background gauge field strength. In particular, for
canonically-normalized gauge bosons, we take the gauge group
generators to include a factor of the corresponding gauge
coupling, $g_a$. Here we use indices $A,B,..$ for the tangent
frame, $M,N,..$ for world indices and lower-case indices to label
gauge-group generators.

Notice that there is an ambiguity in how the mass, $m$, enters
into the above expressions, because the two quantities $X$ and
$m^2$ only enter through their sum: $X + m^2$. As a consequence
there are two ways to use these formulae. On the one hand, one can
lump the physical mass into $X$ and regard the explicit $m$
dependence of eq.~\pref{eqn: heatkernel} as being an infrared
regulator which is taken to zero at the end of the calculation. In
this case only the term with $k = n/2$ survives and the $m$
dependence of $\Sigma$ is completely contained within the $X$
dependence of $a_{n/2}$. Alternatively one can exclude $m^2$ from
$X$, in which case the large-$m$ dependence of $\Sigma$ is
explicit in eq.~\pref{eqn: heatkernel}.

We may use the equivalence of these two points of view to derive
an identity which relates the Gilkey coefficients for $X$ to those
for $X+m^2$. The simplest way to do so is to compute the divergent
part of eq.~\pref{eqn: heatkernel} using the result
$\Gamma(-k-\epsilon) = (-)^k/(k!\epsilon)+\cdots$, for $\epsilon$
an infinitesimal and $k$ a non-negative integer. For odd $n$ this
leads to the old one-loop-finiteness result at one loop in
dimensional regularization \cite{UVFiniteDR}. For even $n$,
comparing the result for the coefficient of $1/\epsilon$ with and
without including $m^2$ in $X$ leads to the following identity:
\beq \label{eqn: gilkeyidentity}
    \tr[a_{n/2}(X+m^2)] = \sum_{k=0}^{n/2} \frac{(-)^{k-n/2}}{(n/2-k)!}
    \tr[m^{n-2k} a_k(X)]
    \,.
\eeq
For instance, for $n=4$ and $n=6$ this reduces to
\bea
    \tr[a_2(X+m^2)] &=& \tr[a_2(X)] - \tr[m^2 a_1(X)] +
    \frac12 \, \tr[m^4 a_0(X)] \nn \\
    \tr[a_3(X+m^2)] &=& \tr[a_3(X)] - \tr[m^2 a_2(X)] + \frac12 \,
    \tr[m^4 a_1(X)] - \frac16 \, \tr[m^6 a_0(X)] \,, \nonumber \\
\eea
which may be verified using the explicit expressions of
eq.~\pref{eqn: gilkey}.

These formulae show that the coefficient of the leading power of
$m$ can be computed by evaluating the first few coefficients,
$a_k$, {\it without} including $m$ explicitly into the quantity
$X$. Provided that the contributions of the would-be Goldstone
bosons and ghosts all share the same $m$ (as we show in detail
below) we may obtain the results for massive fields by summing
appropriate results for massless fields.

We now use this approach to evaluate the first few coefficients,
$\tr(a_k)$ ($k=0,1,2$), in $n$ spacetime dimensions for particles
having spin zero, one-half, one, three-halves and two, as well as
for the rank-two antisymmetric gauge potential which appears in
supergravity models. Although our real interest is to applications
with massive fields, we provide the results for massless fields
which are required as intermediate steps in the calculation.

\subsection{Spin 0}

The lagrangian for a set of $N_0$ real scalar fields,
denoted collectively by $\phi$, is given by
\beq \label{eqn: L_scalar}
    \e \L_0 = -\frac{1}{2} \phi ( -\Box
    + m^2 + \xi R )\, \phi
\eeq
where in general both $m^2$ and $\xi$ are arbitrary constant $N_0
\times N_0$ matrices, and as usual $e = \g$. We here assume for
simplicity that $m^2$ and $\xi$ commute with one another, so a
basis of fields exists for which both are diagonal. A case of
particular interest is the massless, minimally-coupled case, $\xi
= m^2 = 0$, such as would be enforced by a Goldstone-boson
symmetry $\phi \to \phi \, +$ constant. Alternatively, the case
$m^2 = 0$ and
\beq
    \xi = -\frac{(n-2)}{4(n-1)} \, I
\eeq
describes a conformally-invariant coupling for all $N_0$ scalars.

For scalars we have $\Y = -i F^a_{MN} t_a$, where $t_a$ is the
gauge-group generator acting on the scalars of any background
gauge group, under which the scalars are assumed to transform in a
representation $\cR_0$. If this representation contains $N_0$ real
scalars, then we have $\tr(I) = N_0$. For $X = \xi R $ we find
\begin{eqnarray}
\label{eqn: a_scalar}
  \tr_0 (a_0) &=& N_0  \nonumber \\
  \tr_0 (a_1) &=& - \left(\tr \, \xi + \frac{N_0}{6} \right) R  \nn \\
  \tr_0 (a_2) &=& \frac{N_0}{180}  \Bigl[\Riem2 - \Ricci2
  \Bigr] +
  \frac{1}{2} \, \tr \, \left[\left(\xi + \frac{1}{6} \right)^2
  \right] R^2 \nonumber \\
  && \qquad - \frac{1}{6} \, \tr \left( \xi+\frac{1}{5} \right) \Box R
  - \frac{g_a^2}{12} \, C(\cR_0) \f2.
\end{eqnarray}
Here $\tr \xi^k = N_0 \, \xi_0^k$ if all scalars share the same
coupling to $R$ ({\it i.e.} if $\xi = \xi_0 \, I$), and $\tr[t_a
t_b] = g_a^2 \, C(\cR_0) \, \delta_{ab}$, where $C(\cR_0)$ is the
Dynkin index for the scalar representation $\cR_0$. (Our
normalization is such that $C(F)= {k}/{2}$ or $C(A) = Nk$,
respectively, for $k$ fields in the fundamental or adjoint
representations of $SU(N)$.)

\subsection{Spin 1/2}

We take the lagrangian for $N_{1/2}$ spin-half particles to be
\beq \label{eqn: L_spin1/2}
    \e \L_{1/2} =  -\frac{1}{2} \,
    \overline{\psi}(\nott{D} + m )\psi \,,
\eeq
where $\nott{D} = \Gamma^M D_M$ with $\Gamma^M$ denoting the $d
\times d$ Dirac matrices in $n$ dimensions. In $n$ dimensions $d =
2^{[n/2]}$ where $[n/2]$ is the largest integer which is less than
or equal to $n/2$. Since different kinds of spinors are possible
in different spacetime dimensions, it proves useful to define a new
quantity, $\d=2d/\zeta$, where the pre-factor of 2 comes because
we count real fields, and $\zeta=1$, 2, or 4 depending on whether
the spinors in question are Dirac, Majorana or Weyl, or
Majorana-Weyl.\footnote{For a discussion on the allowed spinors in
spacetimes of arbitrary dimension and signature, see for example
\cite{spinors}.}

In order to put the operator $\Delta$ into a form for which
eq.~\pref{eqn: gilkey} applies, we use the fact that (assuming
there are no gauge or Lorentz anomalies) $\log\det( \nott{D} + m
)= \frac{1}{2} \log\det(m^2-\nott{D}^2)$, which implies
\begin{eqnarray}
\label{eqn: sigma_spin1/2}
    i \Sigma_{1/2} &=& \frac{1}{4} \Tr \log \left( m^2 -
    {\nott{D}}^2 \right) \nonumber \\
        &=& \frac{1}{4} \Tr \log \left(-\Box + m^2 - \frac{1}{4} R +
        \frac{i}{2} \Gamma^{AB} F^a_{AB} t_a \right),
\end{eqnarray}
where we use the spin-half result $J_{AB} = - \frac{i}{2}
\Gamma_{AB}$, with $\Gamma_{AB} = \frac{1}{2}[\Gamma_A,\Gamma_B]$.
Thus, we see that eq.~\pref{eqn: gilkey} may be applied if we use
$X =  -\frac{1}{4}R \, I + \frac{i}{2} \Gamma^{AB} F^a_{AB}t_a $,
and divide the overall result by 2 (because of the extra factor of
1/2 in eq. ($\ref{eqn: sigma_spin1/2}$) relative to eq.~\pref{eqn:
gamma1}). Here $I$ denotes the $ \cN_{1/2} \times \cN_{1/2}$ unit
matrix, with $\cN_{1/2} = N_{1/2} \d$.

Using eq. ($\ref{eqn: Y}$), we find in this way
\beq
    \tr \left( \y2 \right) = - \d \, g^2_a \, C(\cR_{1/2}) \f2
    -\frac{1}{8} \cN_{1/2} \Riem2 \,.
\eeq
This leads to the following values for $a_k$:
\begin{eqnarray}
    \label{eqn: a_spin1/2}
    \tr_{1/2}(a_0) &=& \frac{\cN_{1/2}}{2} \nn \\
    \tr_{1/2}(a_1) &=& \frac{\cN_{1/2}}{24} \, R \nn \\
    \tr_{1/2}(a_2) &=& \frac{\cN_{1/2}}{360} \left[ -\frac{7}{8}
    \Riem2 - \Ricci2 + \frac{5}{8}
    \, R^2 + \frac{3}{2} \,\Box R \right]
    \nn \\  &&  \qquad\qquad\qquad\qquad
    + \frac{\d g_a^2}{12} \, C(\cR_{{1}/{2}}) \f2.
\end{eqnarray}

\subsection{Spin 1}

For spins higher than 1/2 the massless and massive cases must be
handled separately, due to the different number of spin states
which are involved in these two cases. This is also related to the
need for gauge symmetries for these higher spins
\cite{GaugeSymHiSpin}, and the possibility of mixing between
higher-spin and lower-spin fields ({\it i.e.} the
Anderson-Higgs-Kibble mechanism). In order to be explicit we first
present the massless case.

\nisubsubsection{Massless Spin 1}

\noindent We start by dividing the total gauge field into a
background component, $A_M^a$, and a fluctuation, $\cA_M^a$,
according to $a^a_M = A^a_M + \cA_M^a$. In terms of these fields
the gauge field strength for the full field, $a_M^a$, becomes
\beq
    f^a_{MN} = F^a_{MN} + D_M \cA_N^a - D_N \cA_M^a +
    {c^a}_{bc} \, \cA^b_M \cA^c_N \,,
\eeq
where $D_M$ is the background covariant derivative built from the
background gauge connection, $A_M^a$, and Christoffel symbol,
and as before $F_{MN}^a$ is the background field-strength tensor.
As usual, the fluctuation, $\cA_M^a$, is chosen to transform in
the adjoint representation under background gauge transformations
--- and so $(t_a)_{bc}=-ic_{abc}$ --- as well as transforming as a
vector under background coordinate transformations.

It is convenient to fix the spin-1 gauge invariance using a
background-covariant gauge-averaging term,
\beq
   \e \L_V^{\, gf} = -\frac{1}{2\xi_1} (D^M \cA_M^a)^2,
\eeq
where $D_M$ denotes the background-covariant derivative built from
the background gauge field and Christoffel symbols. Then expanding
the gauge-field lagrangian,
\beq \label{eqn: L_vector}
    \e ( \L_V + \L_V^{\, gf}) = -\left[ \frac{1}{4} f^a_{MN}
    f_a^{MN} + \frac{1}{2 \, \xi_1} (D^M \cA_M^a)^2 \right] \,,
\eeq
to second order in $\cA^a_M$ and choosing the background-covariant
Feynman gauge ($\xi_1 = 1$), the part of the lagrangian which is
quadratic in the fluctuations, ${\cal L_A}$, becomes
\beq \label{eqn: L_spin1}
    \e {\cal L_A} = -\frac{1}{2} \cA_a^M \Bigl[ -\Box
    g_{MN} \delta^{ab} - Y_{MN}^{ab} + c_c^{\hspace{5pt} ab}
    F^c_{MN} \Bigr] \cA_b^{N} \,,
\eeq
where as before $[D_M,D_N]\cA^{aN} = Y_{MN}^{ab} \, \cA_b^N$.

For a vector field the Lorentz generators are $(J^{AB})_{CD} = -i(
\delta^A_C \delta^B_D - \delta^A_D \delta^B_C)$, and so we see
that $[D_M,D_N] \cA^N_b = R_{MN} \cA^N_b - i \F
({t}_a)_{b}^{\hspace{5pt} c} \cA^{N}_c $. The one-loop
contribution due to vector loops is then given by:
\beqa
    i\Sigma_V &=& -\hf \log \det \Bigl[ \Delta^{M \ a}_{\ \,  N \ b} \Bigr]
    \nn \\
    &=&-\frac{1}{2} \log \det \Bigl[ -\Box \,
    \delta^M_{\ \ N} \delta^a_{\, b} -
    R^M_{\ \, N} \delta^a_{\, b} + 2i F^{c
    M}_{\ \ \ N} ({t}_c)^a_{\ \, b} \Bigr].
\eeqa
We can now see that $X^{M \ \, a}_{\ \ N \ b} = - \eta R^M_{\ \,
N} \delta^a_{b} + 2i ({t}_c)^a_{\ \, b} F^{c M}_{\hspace{11pt}
N}$, where $\eta=\pm1$ is a useful constant to include for later
purposes. For the case considered here we see that $\eta=1$,
whereas when we consider the ghosts associated with spin-2
particles we will find that $\eta=-1$.

For $N_1$ vector fields, we therefore find that $\tr_V(X) = - \eta
N_1 R$ and $\tr_V(X^2)=N_1 \Ricci2 +4g_a^2 \, C(A) \f2$, where
$C(A)$ is the Dynkin index for $N_1$ fields in the adjoint
representation. Similarly, $\tr_V(\y2) = -N_1 \Riem2 - n g_a^2 \,
C(A) \f2$ and $\tr_V(I)= n N_1$. These imply the following results
for vector fields in $n$ spacetime dimensions:
\begin{eqnarray}
 \label{eqn: a_spin1}
 \tr_V(a_0) &=& n N_1 \nn \\
 \tr_V(a_1) &=& \left(\eta-\frac{n}{6}\right)N_1 R  \nn \\
 \tr_V(a_2) &=& \frac{N_1}{360} \Bigl[ (2n-30) \Riem2 + (180-2n)
 \Ricci2  + (5n-60 \eta) R^2  \nn \\
 && \qquad + (60 \eta -12n) \Box R \Bigr]
 + \frac{g_a^2}{12}(24-n) C(A) \f2 \,.
\end{eqnarray}

Since we work in a covariant gauge, to this result must be added
the contributions of the ghosts. For the gauge chosen, the gauge
fixing condition $ f^a = D^M \cA^a_M$ varies under gauge
transformations according to $\delta f^a = \Box \, \epsilon^a$.
Consequently, the lagrangian for the gauge ghosts is
\beq \label{eqn: L_spin1ghost}
    \e \L_{Vgh} = -\, \omega^{*}_a(-\Box \,)
     \omega^a \,,
\eeq
where the $\omega^a$ are complex fields obeying Fermi statistics.
Since this has the same form as the spin zero lagrangian discussed
above (specialized to $\xi = 0$), for the ghosts we may simply
adopt the spin-0 results for the $a_k$, with $N_0 \to N_1$ and
multiplied by an overall factor of $-2$.

Adding the results for vector fields ($\eta=+1$) and ghosts gives
the contribution of physical spin-1 states. Thus, we obtain for
massless spin-1 particles:
\begin{eqnarray}
 \label{eqn: a_PhysicalSpin1}
 \tr_1(a_0) &=& N_1 (n -2) \nn \\
 \tr_1(a_1) &=& \frac{N_1}{6} (8-n) R  \nn \\
 \tr_1(a_2) &=& \frac{N_1}{180} \Bigl[ (n-17) \Riem2 +
 (92-n) \Ricci2 \Bigr] + \frac{N_1}{72} (n-14) R^2  \nn \\
  && \qquad + \frac{N_1}{30} (7-n) \Box R + \frac{g_a^2}{12}(26-n)
  \, C(A) \f2 \,.
\end{eqnarray}

\nisubsubsection{Massive Spin 1}

\noindent If the gauge symmetry is spontaneously broken by the
expectation of a scalar field, $\langle \phi^i \rangle = v^i$,
then the previous discussion is complicated because the part of
the lagrangian quadratic in fluctuations acquires cross terms
between the vector and scalar fields of the form $\cA^a_M t_a
\partial^M \phi$. These terms reflect the physical process
whereby the spin-1 particles acquire masses by absorbing the
scalar fields through the Anderson-Higgs-Kibble mechanism.

In this case the same analysis as above can be performed provided
we average over a more general gauge condition: $f^a = D^M \cA^a_M
+ c \, v \cdot t^a \phi$, with the constant $c$ chosen to remove
the cross terms between $\cA_M^a$ and $\partial_M \phi$. This
simply results in the addition of the same mass matrix $\mu^2$ to
the differential operator $\Delta = -\Box  + X$ for the vector
fields and the ghost fields. This process also results in the
would-be Goldstone bosons ({\it i.e.} the scalar fields which
mixed with the gauge fields) acquiring the same mass matrix,
$\mu^2$ as also appears in the vector-field and ghost actions
\cite{FLSGauge}.

The upshot for massive spin-1 particles is therefore to add the
result for $N_1$ massless spin-1 particles to that of $N_1$
massless scalar fields, with $\xi = 0$. This leads to the
following contributions if the mass $\mu^2$, is not included in
$X$:
\begin{eqnarray}
 \label{eqn: a_MssiveSpin1}
 \tr_{1m}(a_0) &=& N_1 (n -1) \nn \\
 \tr_{1m}(a_1) &=& \frac{N_1}{6} (7-n) R \nn \\
 \tr_{1m}(a_2) &=& \frac{N_1}{180} \Bigl[ (n-16) \Riem2 +
 (91-n) \Ricci2 \Bigr]   + \frac{N_1}{72} (n-13) R^2 \nonumber \\
  && \qquad + \frac{N_1}{30} (6-n) \Box R + \frac{g_a^2}{12}(25-n)
  C(A) \f2 \,.
\end{eqnarray}

\subsection{Antisymmetric Tensors}

We next consider in detail the antisymmetric rank-2 gauge
potential, $B_{MN}$, which appears in supergravity models. As
before we first treat the massless case, and then move on to
massive particles. We also quote the results for massless
antisymmetric tensors of arbitrary rank, as taken from
ref.~\cite{Tseytlin}.

\nisubsubsection{Massless Antisymmetric Tensors}

\noindent The appropriate lagrangian for this field is
\beq
    \e \L_B = -\frac{1}{12} \,  H_{MNP} H^{MNP},
\eeq
where $H_{MNP} = D_{[M}B_{NP]} = 2(D_M B_{NP} +D_{N} B_{PM} + D_{P}
 B_{MN})$,
and to this we add the gauge-fixing term $\e \L_{B}^{\,
gf} = - \frac{1}{2 \xi_{B}} (D_M B^{MN})^2$. Choosing the gauge
parameter to be $\xi_{B} = 1/4$, we obtain the lagrangian
\beq
    \e( \L_B + \L_{B}^{\, gf}) = - B_{MN} \Bigl( -\Box \,
    \overline{\delta}^{MN}_{PQ} + 2 R_{P \ \, Q}^{\ \,
    M \ N} - 2 R^M_{\ \, P} \, \delta^N_{Q} \Bigr) B^{PQ}.
\eeq
Here, $\overline{\delta}^{MN}_{PQ} = \frac{1}{2}(\delta^M_{P}
\delta^N_{Q} - \delta^M_{Q} \delta^N_{P})$ is the appropriate
identity matrix for a rank-2 antisymmetric tensor. The
differential operator which possesses the correct symmetries for
this field is thus seen to be
\beq
   \label{B Delta}
   \Delta^{MN}_{\ \ \ \ PQ} = -\Box \, \overline{\delta}^{MN}_{PQ} +
   (R^{M \hspace{3pt} N}_{\ \, P \ \,  Q} -
   R^{N \hspace{3pt} M}_{\ \, P \ Q}) - \frac12
   (R^M_P \delta^N_Q - R^N_P \delta^M_Q + R^N_Q \delta^M_P -
   R^M_Q \delta^N_P),
\eeq
and so
\beq
   {X^{MN}}_{PQ} = (R^{M \hspace{3pt} N}_{\ \, P \ \,  Q} -
   R^{N \hspace{3pt} M}_{\ \, P \ \, Q}) - \frac12
   (R^M_P \delta^N_Q - R^N_P \delta^M_Q + R^N_Q \delta^M_P -
   R^M_Q \delta^N_P).
\eeq
Similarly $Y_{MN}$ is given by
\beq
  \label{r2 Ymn}
  (Y_{MN})^{AB}_{\ \ \ CD} = \hf ( R^A_{\ \, CMN} \delta^B_D \mp
  R^A_{\ \, DMN} \delta^B_C + R^B_{\ \, DMN} \delta^A_C \mp R^B_{\ \, CMN}
  \delta^A_D ),
\eeq
where for later convenience we also give here the result (bottom
sign) for the rank-2 symmetric tensor field.

Using these expressions for $X$ and $Y_{MN}$, and taking there to
be $N_a$ such antisymmetric gauge potentials, we obtain
\begin{eqnarray}
  \tr_B(X)    &=& N_a (2-n) R \nonumber \\
  \tr_B(X^2)  &=& N_a \Bigl[ \Riem2 + (n-6)\Ricci2 + R^2  \Bigr] \nn \\
  \tr_B(\y2) &=& N_a (2-n) \Riem2 \, ,
\end{eqnarray}
and so are led to the following results for $\tr(a_k)$:
\begin{eqnarray}
   \tr_B(a_0) &=& \frac{N_a}{2} n(n-1) \nn \\
   \tr_B(a_1) &=& -\frac{N_a}{12} (n^2-13n+24)R \nn \\
   \tr_B(a_2) &=&  N_a \left[ \frac{1}{360}(16-n)(15-n) \Riem2
   \nn \right. \\
    && \qquad \left. -
   \frac{1}{360} (n^2-181n+1080) \Ricci2
    + \frac{1}{144}(n^2-25n+120) R^2 \right. \nn \\
    && \qquad \left. -
    \frac{1}{60}(n^2-11n+20) \Box R \right].
\end{eqnarray}

To these expressions must be added the contributions of the
ghosts. The antisymmetric tensor gauge transformations are $\delta
B_{MN} = D_M \Lambda_N -D_N \Lambda_M$, where $\Lambda_M$ is
itself only defined up to a gauge transformation: $\Lambda_M
\rightarrow \Lambda_M + D_M \Phi$. We therefore average over the
secondary gauge-fixing condition $f = D_M \Phi^M$, where $D_M$ is
the appropriate background-covariant derivative. Introducing
ghosts and ghost-for-ghosts for these symmetries, we acquire the
ghost counting of ref.~\cite{KalbRamondCounting}, which states
that each initial tensor gauge potential gives rise to a complex,
fermionic vector ghost, $\omega^M$, and three real, scalar,
bosonic ghosts-for-ghosts,\footnote{The reason we do not obtain
four scalar ghosts, as a naive ghost counting would imply, has to
do with the fact the gauge-fixing function $G_N = D^M B_{MN}$
satisfies the constraint $D^N G_N = 0$. A more detailed discussion
of this point can be found in \cite{KalbRamondCounting}.}
$\phi^i$. Their lagrangians are given by
\begin{eqnarray}
  \e \L_{BVgh} &=& -\omega_M^{*}(-\Box \delta^M_N
  - R^M_N) \omega^N, \nonumber \\
  \e \L_{BSgh} &=& -\hf \phi_i (-\Box) \phi^i\,.
\end{eqnarray}
The contributions of the vector ghosts to $a_k$ is therefore
obtained by replacing $N_1 \to -2N_a$ in the result given above
for vector fields (with $\eta = +1$). Similarly, the scalar ghosts
are obtained from the spin-0 result quoted above, with the
replacements $N_0 \to 3 N_a$ and $\xi \to 0$.

Summing the contribution of the rank-2 tensor and its ghosts leads
to the following expression for the physical massless particles
associated with these antisymmetric tensor fields:
\begin{eqnarray}
 \tr_{a}(a_0) &=& \frac{N_a}{2}(n-2)(n-3) \nn \\
 \tr_{a}(a_1) &=& -\frac{N_a}{12}(n^2-17n+54)R \nn \\
 \tr_{a}(a_2) &=& \frac{N_a}{360} \Bigl[ (n^2-35n+306)\Riem2 -
 (n^2-185n+1446)\Ricci2 \Bigr] \nn \\
         & & \qquad + \frac{N_a}{144}(n^2-29n+174)R^2 -
         \frac{N_a}{60}(n^2-15n+46) \Box R \,.
\end{eqnarray}

\nisubsubsection{Massive Particles}

The particles associated with antisymmetric tensor fields can also
acquire mass through an Anderson-Higgs-Kibble mechanism, in which
the antisymmetric tensor particle `eats' an ordinary gauge field,
$V_M$ \cite{HiggsForm}. As before, a modification of the gauge
choice is required in this case in order not to have mixing terms
of the form $B^{MN} \partial_M V_N$. As we now show, the
contribution of each massive tensor particle is given by adding
the above result for a massless particle to the result for an
$\eta = +1$ massless abelian
--- so with $C(A) = 0$ --- gauge field (including its ghosts).

To demonstrate this explicitly, we start with the lagrangian
\be
  \label{Bmassive}
  \e \L_{mB} = -\frac{1}{12} H_{MNP}H^{MNP} - \frac{1}{4}(V_{MN} -
  2 m \, B_{MN})^2,
\ee
where $V_{MN} = D_M V_N - D_N V_M$ is the field strength of the
abelian gauge field $V_M$, and $m$ is a constant with dimensions
of mass. This lagrangian is invariant under
\bea \label{Bmassivegauge}
    \delta B_{MN} &=& D_M \Lambda_N - D_N \Lambda_M \nn \\
    \delta V_M &=& 2 m \, \Lambda_M + 2 \, \partial_M \sigma,
\eea
where $\Lambda_M$ and $\sigma$ are arbitrary gauge parameters. As
in the massless case, this set of gauge transformations is itself
invariant under a gauge transformation,
\bea \label{GhostGaugeFreedom}
    \delta \Lambda_M &=& \partial_M \epsilon \nn \\
    \delta \sigma &=& -m \, \epsilon \,,
\eea
for an arbitrary function $\epsilon$. As before,
therefore, we find that the ghosts themselves have ghosts. Note that in
the limit $m \rightarrow 0$ the lagrangian
decouples into the lagrangian for a massless antisymmetric tensor
and a massless vector.

To fix the two gauge freedoms in eq.~\pref{Bmassivegauge}, and to
remove unwanted mixing terms, we add to the lagrangian the
gauge-fixing term
\be
   \label{massiveB gf}
   \e \L_{mB}^{\, gf} = -2\left(  D^M B_{MN}
   - \frac{m}{2} \, V_N \right)^2  - \hf
   \left( D^M V_M \right)^2.
\ee
After adding this term to eq.~\pref{Bmassive} we find
\be
   \e ( \L + \L_{mB}^{\, gf}) = - B_{MN} ( \Delta_{\ \ \ \ PQ}^{MN} +
   m^2 \, \bar{\delta}_{PQ}^{MN}) B^{PQ} - \hf V_M (\Delta_N^M +
   m^2 \, \delta_N^M) V^N,
\ee
where $\Delta$ is the differential operator appropriate for the
field it operates on; specifically, $\Delta^M_N = -\Box
\,\delta^M_N - R_N^M$, and $\Delta_{\ \ \ \ PQ}^{MN}$ is given by
eq.~\pref{B Delta}.

The lagrangian for the ghosts is obtained by varying the
gauge-fixing conditions appearing in eq.~\pref{massiveB gf}, and
we thus find
\bea
    \L_{mBgh} &=& - \xi_N^{*} (-\Box \delta^N_M + D_M D^N
    +  m^2 \delta_M^N) \, \xi^M
    -  \omega^{*} (-\Box) \, \omega  \nn \\
    && -  m \, \xi_M^* D^M \omega
    + m \, \omega^{*} D_M \xi^M \,.
\eea
Here, $\xi_M$ and $\omega$ are the ghost fields associated with
$\Lambda_M$ and $\sigma$, respectively. To fix the gauge freedom
implied by eq.~\pref{GhostGaugeFreedom}, we add  to the ghost lagrangian
the term
\be
    \L_{mBgh}^{\, gf} = - {(D_M \xi^{M} + m \omega^{})}^*
    ( D_N \xi^{N} + m \omega )
\ee
and so we find
\be
    \L_{mBgh} + \L_{mBgh}^{\, gf} = - \xi_N^{*}
    \Bigl[( -\Box + m^2 ) \delta_M^N + R_M^N \Bigr]
    \xi^M - \omega^{*} ( -\Box + m^2 ) \, \omega.
\ee
Notice that the complex scalar ghost, $\omega$, combines with the
vector, $V_N$, to form the field content of a physical massless
spin-1 particle.

The ghosts-for-ghosts lagrangian is similarly obtained, and as in
the massless case we find three bosonic scalar ghosts-for-ghosts,
with lagrangian
\be
    \L_{mBSgh} = -\hf \phi_i (-\Box + m^2) \phi^i \,.
\ee
Except for the presence of mass terms, the lagrangian for a
massive antisymmetric tensor is therefore the sum of a massless
spin-1 lagrangian and a massless antisymmetric tensor lagrangian
(including their ghosts). Thus, in calculating the $a_k$ for a
massive antisymmetric tensor, we simply need to add to the
massless result given in the previous section the result for a
massless spin-1 field. It is important to emphasize that such a
sum --- where we factor all mass terms out of $X$, as described in
the $\S \, \ref{GilkeyDeWitt}$ --- makes sense only because in the
gauge we have chosen all particles share the same mass.

The result of this sum, for massive rank-2 tensor fields in $n$
spacetime dimensions, is
\begin{eqnarray}
 \tr_{am}(a_0) &=& \frac{N_a}{2}(n-2)(n-1) \nn \\
 \tr_{am}(a_1) &=& -\frac{N_a}{12}(n^2-15n+38)R \nn \\
 \tr_{am}(a_2) &=& \frac{N_a}{360}\Bigl[ (n^2-33n+272)\Riem2 -
 (n^2-183n +1262)\Ricci2 \Bigr] \nn \\
 && \qquad + \frac{N_a}{144}(n^2-27n+146)R^2 -
 \frac{N_a}{60}(n^2-13n+32) \Box R \,.
\end{eqnarray}

\nisubsubsection{Higher-Rank Antisymmetric Tensors}

\noindent The result for a higher-rank massless skew-tensor gauge
potential in $n$ dimensions has been worked out in a similar
fashion to the above \cite{Tseytlin}. This leads to the following
results for the first few Gilkey coefficients for a massless
3-form gauge field (for $n > 4$ dimensions), specialized to
Ricci-flat background geometries ($R_{MN} = 0$):
\begin{eqnarray}
 \tr_{3a}(a_0) &=& \frac{N_{3a}}{3!}(n-2)(n-3)(n-4) \nn \\
 \tr_{3a}(a_1) &=& 0 \nn \\
 \tr_{3a}(a_2) &=& \frac{N_{3a}}{1080} (n^3 - 54n^2 + 971 n -
  4164)\Riem2 \,.
\end{eqnarray}
The analogous results for a massless 4-form gauge field (in $n
> 5$ Ricci-flat dimensions) are given by:
\begin{eqnarray}
 \tr_{4a}(a_0) &=& \frac{N_{4a}}{4!}(n-2)(n-3)(n-4)(n-5) \nn \\
 \tr_{4a}(a_1) &=& 0 \nn \\
 \tr_{4a}(a_2) &=& \frac{N_{4a}}{4320}  (n^4 - 74 n^3
 + 2051 n^2 - 18634 n + 52680)\Riem2 \,.\nn \\
\end{eqnarray}

These results for massive 1- and 2-forms suggest a short-cut for
extending our results to the case of a massive $p$-form for
arbitrary $p$, since they show that the Gilkey coefficients for a
massive $p$-form are obtained by summing the contributions of a
massless $(p-1)$-form to that of a massless $p$-form. It can also
be readily seen that the Gilkey coefficients for a massive
spin-1 field are obtained by the replacement $n \rightarrow (n+1)$
in the massless formulae, and similarly for the antisymmetric
2-form. One way to see why this should give the correct result is
to reason as follows. It is clear that (for a Minkowski-space
background) a massless $p$-form in $(n+1)$ dimensions and a
massive $p$-form in $n$ dimensions share the same little group,
$SO(n-1)$, and transform in the same representation of this group.
This connection can also be made more explicit by dimensionally
reducing an $(n+1)$-dimensional massless $p$-form on $S^1$ to
obtain a Kaluza-Klein tower of massive $p$-forms in the
lower-dimensional theory. Each massive field is thereby seen to
contain the spin content of an $n$-dimensional massless $p$- and
$(p-1)$-form. A final check on this reasoning can be had using the
results of ref.~\cite{Tseytlin}, which show that the first few
Gilkey coefficients for a massless $(n+1)$-dimensional $p$-form
--- and hence a massive $n$-dimensionsal $p$-form --- are the same
as the sum of the coefficients for a massless $p$- and
$(p-1)$-form in $n$ dimensions.

\subsection{Spin 3/2}

Before proceeding with spin-3/2 and spin-2 particles, we first
pause to establish a few of our supergravity conventions. Our
starting point is the coupled Einstein/Rarita-Schwinger
system. We take the spin-2 field to be described by the standard
Einstein-Hilbert action, which in our conventions is

\beq
    \label{eqn: L spin2}
    \e \L_{EH} = -\frac{1}{2\kappa^2} R,
\eeq
with $\kappa^2 = 8\pi G_N$. For the moment, we do not include a
cosmological term; the generalization of the massless and massive
spin-3/2 particle to the case of a nonzero cosmological constant
is given in the appendix.

The spin-3/2 particle is described by a vector-spinor field,
$\psi_M$, with a kinetic term given by the lagrangian
\beq
    \label{eqn: L_spin3/2}
    \e \L_{VS} = -\frac{1}{2}  \overline{\psi}_M \Gamma^{MNP}
    D_N \psi_P  \,.
\eeq
As before, we use indices $A,B,..$ for the tangent frame, $M,N,..$
for world indices and lower-case indices to label gauge-group
generators. Conversion between tangent and world indices is
accomplished using the vielbein, ${e_M}^A$. Here,
$\Gamma^{ABC}=\frac{1}{6} [\Gamma^A \Gamma^B \Gamma^C + \cdots]$
and $\Gamma^{AB} = \frac12 [\Gamma^A,\Gamma^B]$ are normalized
completely antisymmetric combinations of gamma matrices.

The covariant derivative appearing in eq.~\pref{eqn: L_spin3/2}
can involve background gauge fields in addition to the Christoffel
connection, but only if the corresponding gauge symmetry does not
commute with supersymmetry. Such transformations are particularly
rich when there is more than one supersymmetry in the problem.
Gravitini cannot carry charges for internal symmetries which
commute with supersymmetry, because for these the gravitino must
share the charge of the graviton, which is neutral under all gauge
transformations.

When there are no gauge fields in $D_M \psi_N$, it is
straightforward to verify that the combination $\L_{VS} + \L_{EH}$
is invariant under the linearized supersymmetry transformations
\beq
    \label{eqn: g_gauge}
    \delta e_M^A = -\frac{\kappa}{4} \, \overline{\psi}_M \Gamma^A
    \epsilon + {\rm c.c.} \,,\qquad
    \delta \psi_M = \frac{1}{\kappa} \, D_M \epsilon \,.
\eeq
When background gauge fields {\it are} present in $D_M \psi_N$,
the combination $\L_{VS} + \L_{EH}$ varies into terms involving
these gauge fields. These terms then cancel against variations of
the gauge-field kinetic terms and with gauge-field-dependent terms
in the gravitino transformation law. This shows that gauge fields
for symmetries which do not commute with supersymmetry are special
in that they are intimately related to the gravitini by
supersymmetry.

\nisubsubsection{Massless Spin 3/2}

\noindent In order to put the spin-3/2 lagrangian into a form for
which the general expressions for the Gilkey coefficients apply,
it is convenient to use the following gauge-averaging term,
\beq \label{eqn: 3/2gf}
    \e L_{VS}^{\, gf} = -\frac{1}{2 \,\xi_{3/2}} \,
    (\overline{\Gamma \cdot \psi})
    \nott{D}(\Gamma \cdot \psi) \,.
\eeq
With this term, and after making the field redefinition $\psi_M
\rightarrow \psi_M + A \Gamma_M \Gamma \cdot \psi$, we find that
the lagrangian simplifies in the desired way when we make the
following choices for $A$ and $\xi_{3/2}$:
\beq
    \label{eqn: spin3/2_gaugechoices}
    A  = \frac{1}{2-n}  \qquad \hbox{and} \qquad
    \frac{1}{\xi_{3/2}} = \frac{2-n}{4} \,.
\eeq
These choices allow the vector-spinor lagrangian to be written as
\beq
    \e (\L_{VS} + \L_{VS}^{\, gf}) = -\frac{1}{2} \, \,
    \overline{\psi}_M \nott{D}
    \psi^M \,,
\eeq
and so give the one-loop contribution
\beq
    i \Sigma = \frac{1}{2} \log \det \Bigl[(\nott{D})^
    A_{\hspace{5pt} B} \Bigr] = \frac{1}{4}\log\det
    \Bigl[(-\nott{D}^2)^A_{\hspace{5pt} B} \Bigr]\,.
\eeq

For a vector-spinor the Lorentz generators are
\beq
    (J_{AB})^C_{\hspace{5pt} D} = -\frac{i}{2} \Gamma_{AB}
    \delta^C_D - i I (\delta^C_A \eta_{BD} - \delta^C_B \eta_{AD}) ,
\eeq
where $I$ is the $\cN_{3/2} \times \cN_{3/2}$ identity matrix,
corresponding to the $\cN_{3/2} = N_{3/2} \, \d$ (unwritten)
non-vector components of $\psi_M$. (Recall $\d =
2^{[n/2]+1}/\zeta$, where $\zeta = 1,2$ and 4 for Dirac, Weyl (or
Majorana) and Majorana-Weyl fermions.) Using the identity
$\nott{D}^2 = \Box + \frac{1}{4} [ \Gamma^M,\Gamma^N ][D_M, D_N]
$, we find
\beq
    [ \nott{D}^2]^A_{\hspace{5pt} B} = \left(\Box + \frac{1}{4} R
    -\frac{i}{2} F^{{a}}_{CD} \Gamma^{CD} \, t_{{a}} \right) \delta^A_B
    - \frac{1}{2} R^A_{\hspace{5pt} BCD} \Gamma^{CD}.
\eeq
For simplicity of notation, we have suppressed writing the various
identity matrices that appear in the above expression. From this
we may read off the expression for $X$, given by \beq
    X^{A}_{\hspace{5pt} B} = \left(- \frac{1}{4}R +\frac{i}{2}
    F^{{a}}_{CD} \Gamma^{CD} \, t_{{a}} \right) \,
    \delta^{A}_{B} + \frac{1}{2}
    R^{A}_{\hspace{5pt} BMN} \Gamma^{MN} .
\eeq
Taking appropriate traces, we obtain the results
\begin{eqnarray}
    \tr_{VS}(X)   &=& - \, \frac{n}{4} \cN_{3/2}  \,  R \nn \\
    \tr_{VS}(X^2) &=& \cN_{3/2} \left[ \frac{n}{16} \, R^2
    + \frac{1}{2}  \Riem2 \right]
    + \frac{n \d g_a^2}{2}\, C(\cR_{3/2})
    F^{{a} \hspace{9pt}}_{MN} F_{a \hspace{9pt}}^{MN} \nn \\
    \tr_{VS}\left(\y2 \right) &=& - \cN_{3/2}
    \left(1 + \frac{n}{8} \right) \Riem2
    - n \d g_a^2\; C(\cR_{3/2}) F^{{a}
    \hspace{9pt}}_{MN} F_{{a} \hspace{9pt}}^{MN}. \nn \\
\end{eqnarray}
$\cR_{3/2}$ denotes, as usual, the Dynkin index for the
representation of the gauge group carried by the spin-3/2 fields.

Combining these results, and remembering to multiply (as for the
spin-1/2 case) eq.~\pref{eqn: gilkey} by an overall factor of
1/2, we find
\begin{eqnarray}
    \label{eqn: a vs}
    \tr_{VS}(a_0) &=& \frac{n}{2} \, \cN_{3/2} \nn \\
    \tr_{VS}(a_1) &=&
    %\left(
    \frac{n }{24} \, \cN_{3/2} R
    %-
    %\frac{1}{2} m_{3/2}^2 \right)
    \nn \\
    \tr_{VS}(a_2) &=& \frac{\cN_{3/2}}{360} \left[
    \left(30-\frac{7n}{8} \right) \Riem2 - n \Ricci2 +
    \frac{5n}{8} R^2 + \frac{3n}{2} \Box R \right] \nn \\
    & & \qquad\qquad \qquad
    + \frac{n \d g_a^2}{12} \, C(\cR_{3/2}) F^{{a} \hspace{9pt}}_{MN}
    F_{a \hspace{9pt}}^{MN}
%    + \frac{n}{24} \cN_{3/2} \left( -m_{3/2}^2 R + 6 \, m_{3/2}^4
%    \right)
    \,.
\end{eqnarray}

We next consider the contribution from the ghost fields. From the
supersymmetry transformation rules, we see that $\delta( \Gamma
\cdot \psi ) = \frac{1}{\kappa}  \nott{D} \, \epsilon$  and so
there are two bosonic, Faddeev-Popov spinor ghosts with the
lagrangian
\beq
    \e \L_{LVFPgh} = -\overline{\omega}^i
    %\left(
    \nott{D}
    %+ \frac{n \,
    %m_{3/2}}{n-2} \right)
    \, \omega_i, \label{eqn: FadeevGhost}
\eeq
where $i=1, \,2$ labels the two ghosts. Since this has the same
form as the spin-1/2 lagrangian used earlier, eq. ($\ref{eqn:
L_spin1/2}$), the Faddeev-Popov ghost result for $\tr[a_k]$ is
obtained by multiplying the massless spin-1/2 result by $-2$.

In addition to the Faddeev-Popov ghosts, there is also a bosonic,
Nielsen-Kallosh ghost \cite{NKGhost} coming from the use of the
operator $\nott{D}$ in the gauge-fixing lagrangian, eq.~\pref{eqn:
3/2gf}. The Nielsen-Kallosh ghost lagrangian is given by
\beq
    \e \L_{LVNKgh} = -\overline{\eta}
    %\left(
    \nott{D}
    %- \frac{n \,
    %m_{3/2}}{n-2} \right)
    \, \eta \label{eqn: NKGhost} \,.
\eeq
This ghost therefore has a contribution to $\tr[a_k]$ given by $-1$
times the massless spin-1/2 result.

Adding the results for the Faddeev-Popov and Nielsen-Kallosh
ghosts to that of the vector-spinor, we obtain the following
results for the contribution to $\tr[a_k]$ by physical massless
spin-3/2 states:
\begin{eqnarray}
    \label{eqn: a spin3/2}
    \tr_{3/2}(a_0) &=& \frac{\cN_{3/2}}{2} (n-3) \nn \\
    \tr_{3/2}(a_1) &=& \frac{\cN_{3/2}}{24}
    %\left(
    (n-3)R
    %- \frac{6n(n^2-7n+4)}{(n-1)(n-2)} \Lambda \right)
    \nn  \\
    \tr_{3/2}(a_2) &=& \frac{\cN_{3/2}}{360} \left[
    \left( 30-\frac{7}{8}(n-3) \right)
    \Riem2 - (n-3) \Ricci2 \right. \nn \\
    && \qquad \left. + \frac{5}{8}(n-3)R^2
     + \frac{3}{2} (n-3) \Box R
    %- \frac{15n(n^2-7n+4)}{2(n-1)(n-2)}
    %\Lambda R  \nn \\
    %& & \qquad \left. +\frac{45n(n^4-11n^3+24n^2-32n+16)}{2(n-1)^2(n-2)^2}
    %\Lambda^2
    \right]
    %\nn \\
    %&& \qquad
    + \frac{\d g_a^2}{12}(n-3) \, C(\cR_{3/2}) \, \f2. \nn \\
\end{eqnarray}

\nisubsubsection{Massive Spin 3/2}

A spin-3/2 state acquires a mass through the existence of an
off-diagonal coupling of the form $\overline{\chi} \, \Gamma \cdot
\psi$ with a spin-1/2 Goldstone fermion state, $\chi$. Choosing a
gauge for which this term vanishes causes the super-Higgs
mechanism to occur, through which the spin-3/2 particle `eats' the
fermion $\chi$. Although $\chi$ vanishes in a unitary gauge, it
remains in the theory in a covariant gauge much as does the
would-be Goldstone boson for the massive spin-1 case.

To show explicitly how this process occurs, we assume that the
part of the fermionic lagrangian which is quadratic in the
fluctuations has the general form\footnote{We follow here the
approach of ref.~\cite{Elise} to identify the form of these
couplings to quadratic order in a model-independent way.}
\bea \label{LmVS}
    \e \L_{mVS} &=& -\overline{\psi}_M \Gamma^{MNP} D_N
    \psi_P - \overline{\chi} \nott{D} \chi -
    \Bigl[ \overline{\psi} \cdot \Gamma( a
    \nott{D} + b) \chi + {\rm c.c.} \Bigr] \nn \\
    && - \left( c \, \overline{\psi}_M
    D^M \chi + {\rm c.c.} \right) - m_{1/2} \,
    \overline{\chi} \chi - \mu_{3/2} \,
    \overline{\psi}_M \psi^M \nn \\
    &&+ m_{3/2} \overline{\psi}_M \Gamma^{MN}
    \psi_N \,,
\eea
where the parameters $a$, $b$, $c$, $m_{1/2}$, $m_{3/2}$, and
$\mu_{\, 3/2}$ are constrained by demanding that the action be
invariant under linearized supersymmetry transformations. For
simplicity we assume these parameters to be real, although in
general some or all of these parameters may be complex, depending
on whether the fermions are Majorana or Weyl in the supergravity
of interest. Requiring invariance under the supersymmetry
transformations
\be \label{susytrans}
    \delta \psi_M = \frac{1}{\kappa} D_M
    \epsilon + \mu \Gamma_M \epsilon \qquad \hbox{and} \qquad
    \delta \chi = f \epsilon \,,
\ee
then imposes the following constraints on the various parameters:
\bea \label{params}
    &&a = c = \mu_{\, 3/2} = 0 \qquad
    b = \kappa f \qquad
    f^2 = (n-1)(n-2) \mu^2 \nn \\
    && m_{1/2} = n \kappa \, \mu \qquad
    m_{3/2} = (n-2) \kappa \, \mu \,.
\eea
This leaves one free parameter --- which we can take to be $\mu$,
$f$, or $b$ --- having the physical interpretation of being the
supersymmetry breaking scale.

With these choices, the variation of the gravitino/goldstino
lagrangian is
\be
    \e \, \delta \L_{mVS} = \frac{1}{2\kappa} \, G^{MN} \overline{\psi}_M
    \Gamma_N \epsilon + {\rm c.c.},
\ee
where $G^{MN} = R^{MN} - \frac12 \, R \, g^{MN}$ is the Einstein
tensor. This term is cancelled in the usual way by the variation
of the Einstein-Hilbert action under the graviton transformation
\be
    \delta e_M^{\hspace{8pt} A} = -\frac{\kappa}{4}
    \, \overline{\psi}_M
    \Gamma^A \epsilon + {\rm c.c.}.
\ee

To this lagrangian we add the gauge-fixing term
\be \label{mVSgf}
    \frac{1}{e} {\mathcal L}_{mVS}^{\, gf} = -\overline{F}(
    \nott{D} + \gamma ) F,
\ee
where
\be
    F = \alpha \Gamma \cdot \psi + \beta \chi.
\ee
The constants $\alpha$, $\beta$, and $\gamma$ are chosen to ensure
that the gauge-fixed lagrangian has the form
\be
  \label{LmVSgf}
  \e (\L_{mVS} + \L_{mVS}^{\, gf}) = - {\overline{\psi}'}_M
  ( \nott{D} + {m'}_{3/2})
  {\psi'}^{M} - \overline{\chi}' (\nott{D} + {m'}_{1/2}) \chi'
%  +({\rm ghost \hspace{5pt} terms}),
\ee
where ${\psi'}_M$ and $\chi'$ are given by
\be \label{redefns}
    \chi' = A \chi + B \Gamma \cdot \psi \qquad \hbox{and}
    \qquad
    {\psi'}_M = \psi_M + C \Gamma_M \Gamma \cdot \psi
    + D \Gamma_M \chi \,,
\ee
where we again take the parameters $A$, $B$, $C$, and $D$ to be
real for simplicity. Note that the transformation of $\psi_M$ is
nonsingular provided $C \neq -1/n$. Using eq. ($\ref{redefns}$) to
evaluate the right-hand side of eq. ($\ref{LmVSgf}$) while using
eqs. ($\ref{LmVS}$), ($\ref{params}$), and ($\ref{mVSgf}$) to
evaluate the left-hand side, leads to the conditions
\bea
    \label{mVS conditions}
    &&A= \left( \frac{n-1}{n-2} \right)^{1/2}, \qquad
    B = C = -\hf, \qquad D=0, \qquad
    {m'}_{3/2}={m'}_{1/2}= (n-2)\kappa \mu, \nn \\
    &&\alpha = -\hf \sqrt{n-1}, \qquad
    \beta = \frac{1}{\sqrt{n-2}}, \qquad
    \gamma = -(n-2)\kappa \mu.
\eea

The ghost action consists of a Nielsen-Kallosh ghost, with
lagrangian
\be \label{LmVSNK}
    \e \L_{mVSNK} = -\overline{\omega} ( \nott{D} + \gamma ) \,\omega,
\ee
as well as two Faddeev-Popov ghosts, with lagrangian
\be \label{LmVSFP}
    \e \L_{mVSFP} = -\overline{\xi}_i \Bigl[ \nott{D} + (n-2)\kappa \mu
    \Bigr] \xi^i \,,
\ee
where $i=1,2$ labels the two ghosts. Dropping the primes, and
defining $m = (n-2)\kappa \mu$, the complete lagrangian,
eqs.~\pref{LmVSgf}, \pref{LmVSNK} and \pref{LmVSFP}, becomes
\be
%    \e (\L_{mVS} + \L_{mVS}^{\, gf})
    \e \L_{m3/2}
    = - \overline{\psi}_M
    ( \nott{D} + m ) {\psi^{M}} -
    \overline{\chi}(\nott{D} + m) \chi - \overline{\omega}
    ( \nott{D} - m ) \omega
    -\overline{\xi}_i ( \nott{D} + m ) \xi^i \,.
\ee

Since the heat-kernel coefficents are even under $m \to -m$, we
see from this that $a_k$ for a massive gravitino are given by the
sum of the corresponding coefficients for a massless gravitino
(including ghosts) plus those of a massless fermion. Summing the
massive spin-1/2 result, eq.~($\ref{eqn: a_spin1/2}$), with the
spin-3/2 result, eq. ($\ref{eqn: a spin3/2}$), we obtain the
following Gilkey coefficients for a massive spin-3/2 particle
\begin{eqnarray}
    \label{eqn: a mspin3/2}
    \tr_{m3/2}(a_0) &=& \frac{\cN_{3/2}}{2} (n-2) \nn \\
    \tr_{m3/2}(a_1) &=& \frac{\cN_{3/2}}{24} (n-2)R    \nn \\
    \tr_{m3/2}(a_2) &=& \frac{\cN_{3/2}}{360} \left[
    \left(30-\frac{7}{8}(n-2) \right)
    \Riem2 - (n-2) \Ricci2 \nonumber \right. \\
    % && \qquad
    && \left. + \frac{5}{8}(n-2)R^2 + \frac{3}{2} (n-2) \Box R
    \right] + \frac{g_a^2}{12}(n-2) \d \, C(R_{3/2}) \f2 \,. \nn  \\
\end{eqnarray}

\subsection{Spin 2}

Finally, we turn to spin-2 particles. In order to maximize the
utility of this section, we do so for the case where the
lagrangian includes a cosmological constant, as is typically true
for non-supersymmetric theories (and for supersymmetric theories
in four dimensions), and so start with the following action
\beq
    \label{eqn: L spin2lambda}
    \e \L_{EH} = -\frac{1}{2\kappa^2} (R-2\Lambda) \,.
\eeq
For situations where $\Lambda$ represents the value of a scalar
potential, $V$, evaluated at the classical background, we see from
the above that $\Lambda = - \kappa^2 \, V$.

Although it is usually true that only a single spin-2 particle is
massless in any given model, we include a parameter $N_2$ which
counts the massive spin-2 states. We do so because there is
typically more than one massive spin-2 state in the models of
interest, typically arising as part of a Kaluza-Klein tower or as
excited string modes.

\nisubsubsection{Massless Spin 2}

\noindent The lagrangian for a massless rank-two symmetric field
is the Einstein-Hilbert action, eq.~($\ref{eqn: L spin2lambda}$).
As usual we write the metric as ${g}_{MN} + 2 \kappa \, h_{MN}$,
where $g_{MN}$ is the background metric and $h_{MN}$ are the
fluctuations. Expanding to quadratic order in these fluctuations,
and adding the gauge-fixing term
\beq
    \frac{1}{e} \L_{EH}^{\, gf} =
    -
%    \frac{1}{4 \kappa^2}
    \left( D^M h_{MN} -
    \frac{1}{2} D_N h^M_{\hspace{7pt} M} \right)^2,
\eeq
we obtain the standard result \cite{Duffcc}
\begin{eqnarray}
    \frac{1}{e} \,
    (\L_{EH} + \L_{EH}^{gf}) &=&  \hf h^{MN} \Bigl[
  \Box h_{MN} + (R-2 \Lambda) h_{MN} - ( h_{MA} R_N^{\hspace{7pt} A}
  + h_{NA} R_M^{\hspace{7pt} A} ) \nn \\
  && \qquad - 2 R_{MANB} h^{AB} \Bigr] + h^{MN} R_{MN} h - \frac{1}{4} h
  \Bigl[ \Box h + (R-2 \Lambda)h \Bigr] \,, \nn \\
\end{eqnarray}
where $h = g^{MN} h_{MN}$.

It is useful to decouple the scalar, $h$, from the traceless
symmetric tensor $\phi_{MN} = h_{MN} - \frac{1}{n} \, h\, g_{MN}$,
in this expression. In terms of these variables the lagrangian is
\begin{eqnarray}
    \label{eqn: spin2_crossterm}
    \frac{1}{e}
    (\L_{EH}+\L_{EH}^{\, gf})
    &=&  \hf \phi^{MN} \Bigl[ \Box \phi_{MN} +(R-2\Lambda) \phi_{MN} -
    \left( \phi_{MA} R_{N}^{\hspace{7pt} A}+ \phi_{NA} R_{M}^{\hspace{7pt} A}
    \right) \nn \\
    && \qquad  - 2 R_{MANB} \phi^{AB} \Bigr] +
    \left(\frac{n-4}{n}\right) \phi^{MN} R_{MN} \, h  \nn \\
    && \qquad - \left(\frac{n-2}{4n}\right) \left[ h \,\Box h +
    \left( \frac{n-4}{n} \right)R \, h^2 - 2\Lambda h^2 \right] \,,
\end{eqnarray}
which shows that these fields decouple if we make the assumption
that the background metric is an Einstein space: $R_{MN} =
\frac{1}{n} \, R\, g_{MN}$. Although it seems restrictive, the
assumption that the background be an Einstein space is actually
reasonably general due to the observation that we lose no
generality if we simplify the one-loop action by using the
classical equations of motion. We are always free to do so because
it is always possible to use a field redefinition to remove any
term in the one-loop action which vanishes when the classical
equations are used \cite{EffTheories}.\footnote{Although it is
always possible to simplify (without loss of generality) the
one-loop action using the classical equations -- see below -- by
excluding things like scalar gradients or background $F_{MN}^a$
this assumption restricts the kinds of solutions to the classical
equations we may entertain.} In the presence of a scalar
potential, $V$, (or cosmological constant, $\Lambda = - \kappa^2
V$), the classical equations may often be written $G_{MN} +
\Lambda g_{MN} = 0$, or $R_{MN} = [2\Lambda/(n-2)] g_{MN}$, and
for any such a configuration our analysis applies.

With this assumption, and canonically normalizing the scalar mode
by taking $\phi = [(n-2)/(2n)]^{1/2} h$, we arrive at the desired
expression:
\begin{eqnarray}
\label{eqn: spin2}
   \frac{1}{e}(\L_{EH}+\L_{EH}^{\,gf}) &=& -\frac{1}{2} \phi^{MN} \Bigl[
   -\Box \, \bar{\delta}^{AB}_{MN} +
   2 R_{M \hspace{4pt} N}^{\hspace{8pt} A
   \hspace{7pt} B} + (R_M^A \delta^B_N
   + R_N^A \delta^B_M) \nn \\
   && - (R-2\Lambda) \bar{\delta}^{AB}_{MN} \Bigr] \phi_{AB}
   -\frac{1}{2}\phi \left[ \Box + \left( \frac{n-4}{n} \right) R
   - 2 \Lambda \right] \phi, \nn \\
\end{eqnarray}
where $\bar{\delta}^{MN}_{AB} = \frac{1}{2}(\delta^M_A \delta^N_B
+ \delta^M_B \delta^N_A)-\frac{1}{n}g^{MN} g_{AB}$ is the unit
matrix appropriate for a traceless symmetric tensor. Notice the
presence of the well-known `wrong' sign for the kinetic term of
the scalar mode $\phi$.

We may now separately compute the contributions of $\phi$ and
$\phi_{MN}$ to the heat-kernel coefficients, $a_k$. From
eq.~\pref{eqn: spin2}, the symmetric traceless differential
operator appropriate for $\phi_{MN}$ is seen to be
\beqa
   \label{Spin2DiffOp}
   \Delta^{MN}_{\ \ \ \ PQ} &=& - \Bigl[ \Box  + (R-2\Lambda)
   \Bigr] \, \bar{\delta}^{MN}_{PQ}
   + (R^{M \hspace{3pt} N}_{\ \, P \ \,  Q} +
   R^{N \hspace{3pt} M}_{\ \, P \ Q})
   - \frac{4}{n}(g_{PQ}R^{MN}+g^{MN}R_{PQ}) \Bigr. \nn \\
   && \Bigl. + \hf (R^M_P \delta^N_Q + R^N_P \delta^M_Q + R^N_Q \delta^M_P +
   R^M_Q \delta^N_P) +\frac{4}{n^2} g^{MN} g_{PQ} R \,,
\eeqa
from which the expression for $X$ can be read off directly. Taking traces of the
relevant quantities, we find
\begin{eqnarray}
  \tr_{symtr}(X)   &=& N_2 \left[ -\frac{1}{2n}(n+2)(n^2-3n+4)R + (n+2)(n-1)
  \Lambda \right] \nn \\
  \tr_{symtr}(X^2) &=& N_2 \left[ 3 \Riem2 +\frac{1}{n}(n^2-2n-32)\Ricci2
  \right. \nn \\
  && \qquad \left. + \frac{1}{2n^2}(n^4-3n^3+16n+32)R^2  \right. \nn \\
  && \qquad \left. - \frac{2}{n}(n+2)(n^2-3n+4)\Lambda R + 2(n+2)(n-1)
  \Lambda^2 \right] \nn \\
  \tr_{symtr}(\y2) &=& -N_2 (n+2) \Riem2.
\end{eqnarray}
Applying eq. ($\ref{eqn:
 gilkey}$), we arrive at the following expressions for $\tr[a_k]$:

\begin{eqnarray}
   \tr_{symtr}(a_0) &=& \frac{N_2}{2}(n+2)(n-1) \nn \\
   \tr_{symtr}(a_1) &=& N_2 \left[ \frac{1}{12n}(n+2)(5n^2-17n+24)R -
   (n+2)(n-1)\Lambda \right] \nn \\
   \tr_{symtr}(a_2) &=&  N_2 \left[ \frac{1}{360}(n^2-29n+478) \Riem2
   \right. \nn \\
   && \qquad - \frac{1}{360n}(n^3-179n^2+358n+5760)\Ricci2 \nn \\
   && \qquad + \frac{1}{144n^2}(25n^4-95n^3+22n^2+480n+1152)R^2 \nn \\
   && \qquad + \frac{1}{30n}(n+2)(2n^2-7n+10) \Box R \nn \\
   && \qquad \left. - \frac{1}{6n}(n+2)(5n^2-17n+24) \Lambda R + (n^2 +n-2)
   \Lambda^2 \right].
\end{eqnarray}

The scalar part of the spin-2 lagrangian is given by
\beq
   \frac{1}{e} \L_{EHs} = \frac{1}{2} \phi \left[ -\Box -
   \left( \frac{n-4}{n} \right) R + 2 \Lambda \right] \, \phi,
\eeq
which, apart from an overall sign, has the same form as
eq.~\pref{eqn: L_scalar} if we make the substitution $\xi R
\rightarrow -\left(\frac{n-4}{n}\right) R + 2 \Lambda$. Since the
overall sign of $\Delta$ contributes a
background-field-independent phase to the action which is
cancelled by a similar contribution from the ghost action (see
below), we may ignore it for the present purposes. With these
comments in mind, we may then use the previous results for spin-0
fields to compute the contribution of $\phi$ to the Gilkey
coefficients, $a_k$.

Finally, we consider the ghosts for the graviton field. Since the
gauge-fixing term is $f_N = D^M h_{MN} -\frac{1}{2} D_N h$, and
the gauge transformations are $\delta h_{MN} = D_M \xi_N + D_N
\xi_M$, we find the transformation property
\beq
    \delta f_N = \Box \, \xi_N - R^M_{\hspace{7pt} N} \xi_M \, ,
\eeq
leading to a complex, fermionic, vector ghost
$\omega_M$ with lagrangian
\beq
    \e \L = -\omega_M^{*} ( -\Box \delta^M_N + R^M_N) \, \omega^N.
\eeq

The contribution of the vector ghost to the Gilkey coefficients is
therefore obtained by multiplying the results found earlier for
the real spin-1 field by an overall factor of $-2$ (and using the
choice $\eta= -1$ in eq.~($\ref{eqn: a_spin1}$)). We thus obtain
the result for the massless graviton in $n$ dimensions (for
background Einstein geometries:\footnote{We drop $\Box R$ in these
expressions with only a tiny loss of generality because $R$ is
necessarily constant for an Einstein space provided $n>2$.}
$R_{MN} = (R/n) \, g_{MN}$)
\begin{eqnarray}
  \tr_2(a_0) &=& \frac{N_2}{2} \, n(n-3) \nn \\
  \tr_2(a_1) &=& N_2\left[ \frac{1}{12}(5n^2-3n+24)R -n(n+1)\Lambda \right]
  \nn \\
  \tr_2(a_2) &=& N_2 \left[ \frac{1}{360}(n^2-33n+540)\Riem2 \right. \nn \\
  && \qquad + \frac{1}{720n}(125n^3-497n^2+486n-1440)R^2 \nn \\
  && \qquad \left. - \frac{n}{6}(5n-7) \Lambda R + n(n+1) \Lambda^2 \right].
\end{eqnarray}

\nisubsubsection{Massive Spin 2}

\noindent We next derive the lagrangian for the massive graviton.
In order to do so we require an expression for the quadratic part
of the massive spin-2 lagrangian, such as might be obtained from a
Kaluza-Klein reduction or as a massive string mode. To keep the
analysis as background-independent as possible, we work with the
most general such action for which the spin-2 state acquires its
mass by mixing with the appropriate Goldstone field, as in the
Anderson-Higgs-Kibble mechanism. We believe that by making this
requirement we capture quite generally the contributions of the
massive spin-2 states which arise in dimensional reduction and as
heavy string modes \cite{Spin2MassiveLagr}.

We start, therefore, with the lagrangian
\bea
    \e \L_{mEH} &=& \e \L_{EH} -
    \frac{1}{4} F_{MN}F^{MN} - a h^{MN} D_M V_N - b V^M D_M h \nn \\
    && - c \, R^{MN} V_M V_N - \hf m_1^2 \, V_M V^M -
%    \frac{1}{2\kappa^2}
    \frac12
    \, m_2^2  \, h_{MN} h^{MN} -
%   \frac{1}{2\kappa^2}
    \frac12
    \, \mu_2^2 \,
    h^2\,,
\eea
where the coefficients $a$, $b$, $c$, $m_1$, $m_2$ and $\mu_2$ are
to be determined by demanding the presence of a non-linearly
realized gauge symmetry (which would correspond to the
diffeomorphisms which do not preserve the background geometry
within the Kaluza-Klein context). $F_{MN}$ is the field strength
$D_M V_N - D_N V_M$, where we take $V_M$ to have the spin content
of a massive spin-1 particle. From the previous sections we see
that this should consist of a specific combination of a massless
vector field, $A_M$, and a would-be Goldstone scalar, $\sigma$.
Accordingly, we make the definition
\be
    V_M = A_M + p \, D_M \sigma \,,
\ee
where the coefficient $p$ is also to be determined in what
follows. Notice that, as defined, any lagrangian built from the
vector field $V_M$ automatically  has the gauge invariance
\be \label{DummyVectorGaugeSymmetry}
    \delta A_M = D_M \epsilon \qquad \hbox{and} \qquad
    \delta \sigma = -\frac{1}{p} \, \epsilon.
\ee
If we desire we may use unitary gauge for this symmetry to remove
$\sigma$ completely from the theory, however this is not a
convenient gauge for our purposes and so in what follows we
instead gauge-fix using a more convenient covariant gauge.

In order to implement the underlying gauge invariance which any
such a spin-2 field must manifest we ask the above lagrangian to
be invariant under the usual spin-2 gauge transformation $\delta
h_{MN} = D_M \xi_N + D_N \xi_M$, supplemented by the
Goldstone-type transformation $\delta V_M = f \xi_M$. This leads
to the following lagrangian\footnote{As a check on this result, we
note that by choosing the gauge where $V_M=0$, we recover the Pauli-Fierz
lagrangian of massive gravity \cite{PauliFierz}. Also, in flat space,
this result agrees (after a suitable field redefinition) with the one
given in \cite{Zinoviev}.}

\bea \label{LmEH}
    \e \L_{mEH} &=& \e \L_{EH} -
    \frac{1}{4} F_{MN}F^{MN} + f \, h^{MN} D_M V_N
    + f \, V^M D_M h \nn \\
    && - R^{MN} V_M V_N - \frac{1}{4}f^2 \,
    h_{MN} h^{MN} + \frac{1}{4}
    f^2 \, h^2 \,,
\eea
corresponding to the choices
\bea
    && a = b = -f, \hspace{5pt} c = 1, \nn \\
    && m_1 = 0, \nn \\
    \hbox{and} \qquad && m_2^2 = -\mu_2^2 =
%    \hf \kappa^2 f^2
    \frac{f^2}{2} \,.
\eea

We now fix the two gauge freedoms of this action in such a way as
to remove the mixings between the various fields having differing
spins. To do so we take for the spin-2 gauge-fixing lagrangian
\be \e \L_{mEH2}^{\, gf} = -
%   \frac{1}{4 \kappa^2}(f_N -2f \kappa^2 V_N)^2,
    \left(f_N - \hf f V_N \right)^2,
\ee
where $f$ is the parameter appearing in the lagrangian
\pref{LmEH}, and as before $f_N$ is defined as $f_N = D^M h_{MN} -
\hf D_N h$. This gauge choice removes the $h^{MN} D_M V_N$ term
from the action and introduces a mass term, $m$, for the vector
field, $V_M$, with $m^2 = \hf f^2$.

To fix the other gauge freedom,
eq.~\pref{DummyVectorGaugeSymmetry}, we add the following
gauge-fixing term
\be
    \e \L_{mEH1}^{\, gf} = -\hf ( D_M A^M + \lambda h + \rho \,
    \sigma)^2,
\ee
with $\lambda$ and $\rho$ being parameters which are chosen to
remove the remaining vector-gravity mixing terms in the quadratic
action. In order to do so we again specialize to the case where
the background spacetime is an Einstein space, which we also take
for simplicity to be a solution to the Einstein equations of the
form $G_{MN} + \Lambda g_{MN} = 0$, or $R_{MN} = [2 \Lambda/(n-2)]
\, g_{MN}$. Using this we see that the removal of cross terms
between $A_M$, $h_{MN}$, and $\sigma$ requires the choices
\be
    \lambda = -\frac{f}{2}, \qquad \hbox{and} \qquad
    \rho = p \, q^2,
\ee
where $q^2$ is defined as
\be
    q^2 = m^2+ \frac{4\Lambda}{n-2} \,,
\ee
since in this case the gauge-fixed lagrangian can be written as
\be
    \L_{mEH} + \L_{mEH1}^{\, gf} + \L_{mEH2}^{\, gf}
    = \L_{mEH0} + \L_{mEH1} + \L_{mEH2}
%    + ({\rm ghost \ \ terms})
    \,,
\ee
with the decoupled lagrangians, $\L_{mEH0}$, $\L_{mEH1}$, and
$\L_{mEH2}$, defined as follows.

$\L_{mEH2}$ denotes the $\phi_{MN}$ lagrangian, which takes the
form
\bea
    \e \L_{mEH2} &=& \frac{1}{e} \L_{EH} -
    f_N f^N - \frac{1}{4} f^2 h_{MN} h^{MN} + \frac{1}{8}
    f^2 h^2 \nn \\
  &=& -\frac{1}{2} \phi_{MN} \left( \Delta^{MN}_{\ \ \ \ PQ} + m^2
  \bar{\delta}^{MN}_{PQ} \right) \phi^{PQ} \nn \\ && +
  \left( \frac{n-2}{4n} \right) h \left(
  - \Box + m^2 + \frac{1}{n}(4-n) R + 2 \Lambda \right) h \nn \\
  && \qquad \qquad + \left( \frac{n-4}{n} \right) R_{MN} \phi^{MN} h \nn \\
  &=& -\hf \phi_{MN} \left( \Delta^{MN}_{\ \ \ \ PQ} +
  m^2 \bar{\delta}^{MN}_{PQ} \right) \phi^{PQ} \nn \\
  &&  \qquad \qquad  +
  \hf \phi \left( - \Box + m^2 + \frac{4\Lambda}{n-2} \right) \phi,
\eea
where $\phi$, $\phi_{MN}$, $\bar{\delta}_{AB}^{MN}$ and
${\Delta^{MN}}_{PQ}$ are as defined above for the massless spin-2
case. The mass $m$ is related to the symmetry-breaking parameter
$f$ by $m^2 = \hf f^2$.

We similarly find the following vector lagrangian, $\L_{mEH1}$:
\bea
    \e \L_{mEH1} &=& -\frac{1}{4} F_{MN} F^{MN}
    -\hf (D^M A_M)^2 - R^{MN} A_M A_N - \hf m^2 A_M A^M \nn \\
    &=& -\hf A_M \Bigl[ (-\Box + m^2)\, \delta^M_N + R^M_N
    \Bigr] A^N,
\eea
where $m$ is the same as for $\phi_{MN}$.

Finally, the part of the quadratic action depending on $\sigma$ is
\be
    \e \L_{mEH0} = \hf \Bigl[ (p \rho) \sigma \Box \sigma -
    (pf) \, \sigma \Box h - (\rho^2) \, \sigma^2 +  (f \rho) h \sigma
    \Bigr] \,,
\ee
which contains terms which mix $\sigma$ and $h$. However, since
$p$ is as yet unspecified we may choose its value to remove these
cross terms. This may be done by choosing
%$p = - \kappa f/(2q^2)$,
$p = - f/(4q^2)$ and making the field redefinition $\tilde{\sigma}
%= \frac{\sqrt{2}m}{4q }( \sigma + \frac{1}{\kappa} h )$, we find
= \frac{\sqrt{2}m}{4q }( \sigma + 2 h )$, after which we find
\bea
    \e \L_{mEH0} &=& -\hf \, \tilde{\sigma} \, \left(
    -\Box + m^2 + \frac{4\Lambda}{n-2}  \right)
    \, \tilde{\sigma} \nn \\ && +
%    \frac{1}{16\kappa^2}\left(\frac{m^2}{q^2}\right)
    \left(\frac{m^2}{4q^2}\right) h \left( -\Box + m^2 +
    \frac{4\Lambda }{n-2} \right) h.
\eea
Notice that in this form the last term in $\L_{mEH0}$ (involving
$h$) has the same form as the last term in $\L_{mEH2}$, and so
these can both be combined into $\L_{mEH2}$ by appropriately
rescaling the scalar $\phi$. Once this is done, and dropping the
tilde on $\sigma$, the remaining term becomes
\bea
    \e \L_{mEH0} &=& -\hf \, \sigma \, \left[
    -\Box + m^2 +
    \frac{4\Lambda }{n-2} \, \right] \, \sigma.
\eea

Finally, the action for the ghosts can be easily calculated from
the gauge-fixing conditions. The spin-2 gauge-fixing term
introduces a complex, fermionic, vector ghost with lagrangian
\be
    \e \L_{mEHVgh} = -\omega_M^* \Bigl[
    (-\Box + m^2) \, \delta^M_N +
    R^M_N \Bigr] \omega^N.
\ee
Similarly, the spin-1 gauge-fixing term introduces a complex
scalar ghost with lagrangian
\be
    \e \L_{mEHSgh} = -\omega^* \left(-\Box + m^2 +
    \frac{4\Lambda }{n-2}
    \right) \omega.
\ee

The complete lagrangian, including all ghosts, for the massive
graviton is thus the sum
\be
%    \L_{mEH} + \L_{mEH1}^{\, gf} + \L_{mEH2}^{\, gf} =
    \L_{m2} = \L_{mEH0}
    + \L_{mEH1} + \L_{mEH2} + \L_{mEHSgh} + \L_{mEHVgh}.
\ee
We are now in a position to assemble the results for $a_k$. To
this end, notice that all fields have been decoupled in the
kinetic terms and all now have the same mass, $m^2 = \hf f^2$. This
allows us to sum the separate contributions to $a_k$ from each of
these fields. It is also interesting to note that the scalar
fields $h$, $\sigma$ and the complex scalar ghost all have
precisely the same lagrangian, and so their net effect is to
completely cancel one another in the one-loop action. Similarly,
the vector boson $A_M$ and the complex vector ghost also share the
same lagrangian, and so for our purposes these two together
contribute the equivalent of one real vector ghost.

In summary, the one-loop divergences for the massive graviton are
given by the sum of the divergences of a symmetric traceless field
and one real vector ghost (for which $\eta=-1$). Thus, we find
\begin{eqnarray}
    \tr_{2m}(a_0) &=& \frac{N_2}{2}(n+1)(n-2) \nn \\
    \tr_{2m}(a_1) &=& N_2 \left[\frac{(6-n)(n+4)(n+1)\Lambda}{6(n-2)} \right] \nn \\
    \tr_{2m}(a_2) &=& N_2 \left[ \frac{1}{360}(n^2-31n+508)\Riem2 \right.
    \nn\\
    && \qquad \left. + \, \frac{(5n^4-7n^3-248n^2-596n-1440)\Lambda^2}{180(n-2)^2}\right] \, \nn \\
\end{eqnarray}
for $n$-dimensional massive gravitons on background metrics
satisfying $G_{MN}+\Lambda g_{MN} =0$.

\section{Supergravity Models}

In supergravity theories the ultraviolet sensitivity of the
low-energy theory is often weaker than in non-supersymmetric
models. This weaker sensitivity arises due to cancellations
between the effects of bosons and fermions in loops. The purpose
of this section is to illustrate the utility of the previous
section's results by using them to exhibit this cancellation
explicitly for supergravities in various dimensions. Some of the
results we obtain --- particularly those for massless particles in
higher-dimensional supergravities --- are computed elsewhere, and
we use the agreement between these earlier calculations and our
results as a check on the validity of our computations.

We proceed by summing the above expressions over the particles
appearing in the appropriate supermultiplets. The result for the
ultraviolet-sensitive part of the one-loop action obtained by
integrating out a supermultiplet is given by
\beq \label{eqn: SGheatkernel}
    \Sigma_{UV} =  \frac{1}{2} \left( \frac{1}{4 \pi} \right)^{n/2}
    \int d^n x \sqrt{-g} \sum_{k=0}^{[n/2]}  \sum_p (-)^{F(p)}
     m_p^{n-2k}\, \Gamma ( k - n/2) \, \tr_p[  a_k ]
    \,,
\eeq
where the sum on $p$ runs over the elements of a supermultiplet.
As is clear from this expression, it is the weighted sum $\sum_p
(-)^{F(p)} \, m_p^{n-2k} \, \tr_p[a_k]$ which is of interest in
supersymmetric theories.

In Minkowski space the strongest suppression of UV sensitivity
arises when supersymmetry is unbroken, in which case all members
of a supermultiplet share the same mass (so that $m_p = m$ for all
$p$). In this case, eq.~\pref{eqn: SGheatkernel} can be written as
\beq
    \label{SigmaSM}
    \Sigma_{UV} =  \frac{1}{2} \left( \frac{1}{4 \pi} \right)^{n/2}
    \int d^n x \sqrt{-g} \sum_{k=0}^{[n/2]} m^{n-2k}  \, \Gamma ( k - n/2) \, \Tr [a_k] \, ,
\eeq
where
\beq
    \Tr[a_k] \equiv \sum_p (-)^{F(p)} \, \tr_p[a_k]
\eeq
is the relevant combination of heat-kernel coefficients for a
supermultiplet. Since $\tr[a_0]$ simply counts the spin states of
the corresponding particle type, the cancellation of the leading
UV sensitivity occurs for a mass-degenerate supermultiplet simply
because each supermultiplet contains equal numbers of bosons and
fermions:
\beq
    \Tr[a_0] = \sum_p (-)^{F(p)} \tr_p[a_0] = N_B - N_F =  0 \,.
\eeq
This ensures the absence of a dependence of the form $m^{n}$ in
$\Sigma_{UV}$.

The story is more complicated when there is a nonzero cosmological
constant, and this is due to the fact that mass itself is more
delicate to define in de Sitter or anti-de Sitter spacetimes. For
Minkowski space mass can be defined for particle states as a
Casimir invariant of the Poincar\'e group, but this definition is
no longer appropriate when $\Lambda$ is nonzero because Poincar\'e
transformations are then not the relevant spacetime isometries.
Rather, for de Sitter space the relevant isometry group in four
dimensions is ${\rm SO}(4,1)$, while the isometries of anti-de
Sitter space fill out the group ${\rm SO}(3,2)$. For these
geometries it only makes sense to inquire about the implications
of unbroken supersymmetry for the anti-de Sitter case. This is
because supersymmetry is always broken in de Sitter spacetime,
whereas there is a supersymmetric generalization of ${\rm
SO}(3,2)$ for which one can find particle supermultiplets which
represent the unbroken supersymmetry.

In our previous calculations of the Gilkey coefficients we have
defined $m^2$ to be that piece in the operator $(-\Box + X)$ which
is a constant for {\it arbitrary} background fields.\nopagebreak
\footnote{This statement requires appropriate modification in the
case of spin 2, where we include a cosmological constant term in
the Lagrangian.} We nevertheless must still grapple with the above
ambiguities as to the meaning of mass in de Sitter and anti-de
Sitter spacetimes, due to the freedom of absorbing into $m^2$
contributions coming from the background curvature for
constant-curvature spacetimes. One can try to restrict this
freedom by demanding masslessness to correspond to conformal
invariance or (for higher-spin fields) to unbroken gauge
invariance, bearing in mind that these choices need not imply
propagation along the light cone \cite{Deser}.

The upshot of this discussion is that it need not be true that all
of the particles within a supermultiplet share the same mass even
when working about a supersymmetric AdS background. In such cases
one cannot pull a common mass out of the sum over particles within
a supermultiplet, as was done in going from eq.~\pref{eqn:
SGheatkernel} to eq.~\pref{SigmaSM}.

To see this concretely, consider the specific example of a
Wess-Zumino multiplet in $n=4$ spacetime dimensions expanded about
a supersymmetric AdS background. Such a multiplet consists of a
scalar, pseudoscalar, and spinor field: $(S,P,\chi)$, and taking
the scalar and pseudoscalar to have a conformal coupling
parameter, $\xi=-1/6$, their mass terms can be written as $m_S^2 =
m^2 - \delta m^2$, $m_P^2 = m^2 + \delta m^2$ and $m_{\chi}^2 =
m^2$. Unbroken supersymmetry implies that these mass terms are
related to one another by $m^2=\mu^2 \Lambda/12$ and $\delta m^2 =
\mu \Lambda/6$, where $\Lambda$ is the AdS cosmological constant
(which is positive in our conventions) and $\mu$ is a
dimensionless parameter which classifies the massive
supersymmetric particle representations. In this case, we find
\bea
   \label{WessZumino}
   \sum_{p} (-)^{F(p)} m_p^4 \, \tr_p[  a_0 ] &=&
   m_S^4  \, \tr_S[a_0] + m_P^4 \,  \tr_P[a_0]
   - m_{\chi}^4 \, \tr_{\chi}[a_0] = 2\, \delta m^4 = \frac{\mu^2 \Lambda^2}{18} \nonumber \\
   \sum_{p} (-)^{F(p)} m_p^2 \, \tr_p[  a_1 ] &=&
   m_S^2  \, \tr_S[a_1] + m_P^2 \,  \tr_P[a_1]
   - m_{\chi}^2 \, \tr_{\chi}[a_1] =- \frac{2 \, m^2 \,
   \Lambda}{3} =-\frac{\mu^2 \Lambda^2}{18} \nonumber \\
   \sum_{p} (-)^{F(p)} m_p^0 \, \tr_p[  a_2 ] &=&
   \tr_S[a_2] + \tr_P[a_2] - \tr_{\chi}[a_2] =
   \frac{R_{MNPQ}^2}{48}-\frac{\Lambda^2}{9}.
\eea

The above complication keeps us from quoting general expressions
for the sum of the Gilkey coefficients over arbitrary
supermultiplets in general dimensions, since for AdS backgrounds
these must be computed with the specific dependence of the
relevant masses on $\Lambda$. Notice however that last expression
in eq.~\pref{WessZumino} contains no dependence on the individual
particle masses (since $\tr_p[a_2]$ is multiplied by $m_p^0=1$).
Terms which are only present in the mass invariant piece of
$\Sigma_{UV}$, such as $R^2_{MNPQ}$ and $F_{MN}^2$, can be
calculated once and for all in a model-independent way because
their coefficients do not depend on the details of the particle
masses involved. This we do in Tables~\pref{4Dtable-sugra} and
\pref{4Dtable-sugramassive} for various 4D supermultiplets. As can
be seen from the above example, however, calculating the complete
answer for $\Sigma_{UV}$ is not difficult once the individual
particle masses are known. Similar considerations hold for
dimensions other than four, with some terms in $\Sigma_{UV}$ being
mass independent and others requiring more detailed knowledge of
the particle spectrum about a given background.

\nisubsubsection{Equations of Motion}

\noindent In the remainder of this section we use the previous
results to compute the statistics-weighted sum of $\tr[a_1]$ and
$\tr[a_2]$ over the particle content obtained by linearizing
various supergravity theories about different solutions to their
classical field equations. To this end we must evaluate results of
the previous section at the solutions to the relevant field
equations,\footnote{Recall that we are always free to use the
classical equations of motion to simplify any one-loop quantity
(like $\Sigma_{UV}$), because any one-loop term which vanishes
with the classical field equations may be removed from $\Sigma$ by
performing an appropriate field redefinition
\cite{ETbooks,EffTheories}.} and sum over the relevant particle
content describing these fluctuations.

The field equations for a very broad class of supergravities
become reasonably simple once restricted only to background
metrics, gauge fields and the scalar dilaton. These equations may
be derived from the action
\beqa
    S &=& - \int d^nx \; \sqrt{-g} \left[ \frac12 \, R
    + \frac12 \, \partial_M \phi
    \, \partial^M \phi + V(\phi) \right. \nn\\
    && \qquad \qquad \qquad \qquad \left. + \frac14 \,
    e^{\lambda\phi} \, \f2 + \frac{1}{2r!} \, e^{\beta \phi}
    \, H_{M_1..M_r}
    H^{M_1..M_r} \right] \,,
\eeqa
where $\lambda$ and $\beta$ are dimension- and
supergravity-dependent numbers and $V$ is a dimension- and
supergravity-dependent potential for the dilaton $\phi$. Notice
that we use units here for which Newton's constant satisfies
$\kappa = 1$.

The simplest class of solutions to these equations are those for
which the gauge fields vanish, $F^a_{MN} = 0$, and the dilaton is
constant, $\partial_M \phi = 0$, at a value for which $V'=0$.
(More general solutions having nonzero background gauge fields,
$F^a_{MN}$, are also possible and usually --- but not always ---
require a non-constant background dilaton configuration as well:
$\partial_M \phi \ne 0$.) In this case the field equations require
the metric to be an Einstein space, $G_{MN} + \Lambda \, g_{MN} =
0$, or
\beq
    R_{MN} =  \left( \frac{2 \Lambda}{n-2} \right) \, g_{MN} \,,
\eeq
where $\Lambda = - V$, evaluated at the vacuum configuration.

\subsection{11D Example}

Eleven-dimensional supergravity has a particularly simple field
content, consisting of a vielbein (or metric), a gravitino, and an
antisymmetric 3-form, and so provides a simple starting example.
Our purpose in this example is to compare with the known results
of ref.~\cite{Tseytlin} as a check on our
calculations.\footnote{For the case of the graviton, the terms we
find proportional to the Ricci scalar appear to differ with those
of \cite{Tseytlin}. However there is no discrepancy once we
specialize to solutions of the equations of motion because their
analysis assumes that $\Lambda=0$, and so their classical
equations of motion require $R=0$.} The contributions to some of
the Gilkey coefficients specialized to 11 dimensions are listed in
Table \pref{11Dtable}.

%Since we have not explicitly worked through the case of a 3-form
%gauge potential, we instead make use of the results of
%ref.~\cite{Tseytlin}, who have performed the calculation for a
%$p$-form in $n$ dimensions.

\begin{table}
\centerline{
\begin{tabular}{|l||c||c||c|c|}
\hline&\multicolumn{1}{c||}{$(-)^F$\tr($a_0$)}&\multicolumn{1}{c||}
{$(-)^F$\tr($a_1$)}
&\multicolumn{2}{c|}{$(-)^F$\tr($a_2$)}\\
\cline{2-5} & 1 & $\frac{1}{3}R$ & $\frac{1}{180}\altRiem2$ &
$\frac{1}{495}R^2$  \\
\hline\hline
a/s 3-form         & 84     & 21    & 219    & 84    \\
gravitino (M)      & $-128$ & $-32$ & $-368$ & $-188$  \\
graviton           & 44     & 149   & 149    & 6884    \\
\hline
\end{tabular}}
\caption{Gilkey coefficients for massless states in 11D, using
$R_{\scriptscriptstyle MN}=(R/n) g_{\scriptscriptstyle MN}$.} \label{11Dtable}
\end{table}

Because the theory has equal numbers of bosons and fermions, we
have $\Tr (a_0) = 0$. Because the background metric is Ricci flat,
it also follows that $\Tr (a_1) = 0$ and $\Tr(a_2) \propto
\Riem2$. Summing the coefficients in Table \pref{11Dtable} then shows that
\beq
    \Tr_{11D}(a_2) = \sum_p (-)^{F(p)} \, \tr_p(a_2) =
    \frac{1}{180} \Bigl[ 219 - 368 + 149 \Bigr] \altRiem2 =0 \,,
\eeq
in agreement with ref.~\cite{Tseytlin}.

The same result can also be obtained for geometries of the form
${\cal M}_6 \times T_5$ without having to use expressions for the
contribution of a 3-form field, simply by truncating the 11D
theory to 6D, such as would be obtained for the massless
Kaluza-Klein spectrum by dimensionally reducing on a 5-torus
\cite{Tseytlin}. The 6D spectrum obtained in this way consists of:
1 graviton, 4 Weyl gravitini, 5 2-form potentials, 16 (1-form)
gauge fields, 20 Weyl fermions, and 25 scalars which we take to be
minimally coupled. Since in 6 dimensions a 3-form potential is
dual to a 1-form, the entire dimensionally-reduced field content
can be handled using the expressions given above. Summing the 6D
results --- given explicitly in Table \pref{table 6D_Massless1}
below --- for this field content, and specializing to the case of
a Ricci-flat 6D metric (with all gauge field strengths vanishing,
$F_{\mu\nu} = 0$), again gives the results $\Tr(a_0) = \Tr(a_1) =
\Tr(a_2) = 0$.

\subsection{10D Examples}

The supergravities of interest in 10 dimensions are those which
arise as the low-energy limits of heterotic, Type I, Type IIA and
Type IIB string theories. Since results for the Gilkey
coefficients are known for each of these, we briefly
consider them in turn. For convenience, the specialization of the
previous sections' formulae to the case $n=10$ is given in Table
\pref{10Dtable}. (This table is also specialized to the choices
$C({\cal R}_0) = C({\cal R}_{3/2}) = 0$, as is appropriate for
these 10D supergravities.)

\begin{table}
\centerline{
\begin{tabular}{|l||c||c||c|c|}
\hline&\multicolumn{1}{c||}{$(-)^F$\tr($a_0$)}&\multicolumn{1}{c||}{$(-)^F$\tr($a_1$)}
&\multicolumn{2}{c|}{$(-)^F$\tr($a_2$)}\\
\cline{2-5}          & 1 & $\frac{1}{6} R$  & $\frac{1}{180}\altRiem2$   & $\frac{1}{12}g_a^2\altf2$ \\
\hline\hline
spin zero ($\xi=0$)  & 1     & $-1$  & 1      & $-$  \\
spin half (M-W)      & $-8$  & $-4$  & 7      & $-16 C({\cal R}_{1/2})$ \\
spin one             & 8     & $-2$  & $-7$   & 16 $C(A)$  \\
a/s 2-form           & 28    & 8     & 28     & $-$ \\
a/s 3-form           & 56    & 34    & 191    & $-$ \\
a/s 4-form           & 70    & 50    & 310    & $-$ \\
gravitino (M-W)      & $-56$ & $-28$ & $-191$ & $-$ \\
graviton             & 35    & 247   & 155    & $-$ \\
\hline
\end{tabular}}
\caption{Gilkey coefficients for massless states in 10D. Terms in
$a_2$ involving only the Ricci tensor or Ricci scalar are not
explicitly displayed. Hyphens indicate quantities which do not
arise and so are not tabulated.} \label{10Dtable}
\end{table}

\nisubsubsection{Type IIA and IIB Theories}

\noindent The field content of the Type IIA theory is given by the
metric, $g_{MN}$, two Majorana-Weyl gravitini having opposite
chiralities, $\psi_M^r$, a 3-form gauge potential, $C_{MNP}$, a
2-form potential, $B_{MN}$, a gauge potential, $C_M$, two
Majorana-Weyl dilatini (with opposite chiralities), $\chi^r$, plus
a dilaton, $\phi$. The dilaton potential vanishes, $\Lambda = -
V=0$.

Summing the contributions of each field to the Gilkey
coefficients, and evaluating at Ricci-flat metrics with vanishing
gauge potentials again gives the result $\Tr(a_0) = \Tr(a_1) = 0$
and
\beq
    \Tr_{IIA}(a_2) = \frac{1}{180} \Bigl[155 - 2(191)
    + 191 + 28 -7 + 2(7) + 1
    \Bigr] \altRiem2 = 0 \,.
\eeq
This may also be understood using the vanishing of these
quantities in 11 dimensions because the Type IIA theory can be
obtained by dimensionally reducing the 11D theory on a circle.

\begin{table}
\centerline{
\begin{tabular}{|l||c||c||c|c|}
\hline&\multicolumn{1}{c||}{$(-)^F$\tr($a_0$)}&\multicolumn{1}{c||}{$(-)^F$\tr($a_1$)}
&\multicolumn{2}{c|}{$(-)^F$\tr($a_2$)}\\
\cline{2-5} & 1 & $\frac{1}{30} R$ & $\frac{1}{180}\altRiem2$ & $\frac{1}{12}g_a^2 \altf2$ \\
\hline\hline
spin zero ($\xi=0$)  & 1      & $-5$   & 1      & $- C({\cal R}_{0}) $  \\
spin one-half (M)    & $-16$  & $-40$  & 14     & $-32 \,C({\cal R}_{1/2})$ \\
spin one             & 9      & $-15$  & $-6$   & $15 \,C(A)$  \\
a/s 2-form           & 36     & 30     & 21     & $-$ \\
a/s 3-form           & 84     & 210    & 219    & $-$ \\
a/s 4-form           & 126    & 420    & 501    & $-$ \\
gravitino (M)        & $-128$ & $-320$ & $-368$ & $-256 \, C({\cal R}_{3/2})$ \\
graviton             & 44     & 1142   & 149    & $-$ \\
\hline
\end{tabular}}
\caption{Gilkey coefficients for massive states in 10D. Terms in
$a_2$ involving only the Ricci tensor or Ricci scalar are not
explicitly displayed. Hyphens indicate quantities which do not
arise and so are not tabulated.} \label{10DtableM}
\end{table}

The field content of the Type IIB theory is obtained from the Type
IIA theory by giving the fermions the same -- rather than opposite
-- chirality and by replacing the 1- and 3-form potentials by a
scalar (0-form), $C$, a 2-form, $C_{MN}$, and a self-dual 4-form,
$C_{MNPQ}$. For this theory the dilaton potential again vanishes
so $\Lambda = -V = 0$. The statistics-weighted sum of the Gilkey
coefficients $a_0$, $a_1$ and $a_2$ again vanishes for this field
content, as may be seen since
\beq
    \Tr_{IIB}(a_2) = \frac{1}{180} \left[155 - 2(191)
     + \frac12(310) + 2(28) + 2(7) + 2(1)
    \right] \altRiem2 = 0 \,.
\eeq
Part of this result can again be understood in a different way,
since the Type IIA and IIB supergravities produce the same theory
when dimensionally reduced on a 2-torus to 9 dimensions. Since we
know from the above that the Type IIA theory gives $\Tr(a_0) =
\Tr(a_1) = \Tr(a_2) = 0$ for this kind of compactification, it
follows that these quantities must also vanish for Type IIB
theories when evaluated on a 9-dimensional Ricci-flat background.

\nisubsubsection{Heterotic and Type I Theories}

\noindent The field content of the Type I and heterotic theories
consist of a 10D $N=1$ supergravity multiplet coupled to a 10D
super-Yang-Mills multiplet for the gauge groups $E_8 \times E_8$
or $SO(32)$, both of which are 496-dimensional.

The $N=1$ supergravity multiplet in 10D consists of: one graviton
$g_{MN}$, one Majorana-Weyl gravitino $\psi_M$, one 2-form
potential $B_{MN}$, one Majorana-Weyl spin-1/2 fermion $\chi$ and
a scalar dilaton $\phi$. For Type I and heterotic models the 10D
gauge multiplet consists of $N_A$ gauge fields $A^a_M$ and $N_A$
Majorana-Weyl spinors $\lambda^a$, where $N_A = 496$ is the
dimension of the gauge group. These supergravities have vanishing
dilaton potential, $V = \Lambda = 0$, but are distinguished from
one another by the value of the gauge-dilaton coupling, which is
given by $\lambda = -4/(n-2) = -1/2$ for the heterotic theory, or
$\lambda = (n-6)/(n-2) = +1/2$ for the Type I theory.

\begin{table}
\centerline{
\begin{tabular}{|l||c||c||c|c|c||c|}
\hline&\multicolumn{1}{c||}{$(-)^F$\tr($a_0$)}&\multicolumn{1}{c||}{$(-)^F$\tr($a_1$)}&\multicolumn{3}{c||}{$(-)^F$\tr($a_2$)}&\multicolumn{1}{c|}{$(-)^F$\tr($a_2$)$|_{ms}$ }\\
\cline{2-7}& 1 & $\frac{1}{10}\Lambda$ & $\frac{1}{360} \altRiem2$ & $\frac{1}{600}\Lambda^2$ & $\frac{1}{12}g^2_a \altf2$ & $\frac{1}{25}\Lambda^2$ \\
\hline\hline
spin zero ($\xi=0$)   &  1 & $-5$  & 2      &  70     & $-C({\cal R}_0)$        & 3        \\
spin zero ($\xi=-1/5$)&  1 & 1     & 2      & $-2$    & $-C({\cal R}_0)$        & 0        \\
spin one-half (W)     &$-4$& $-10$ & 7      & $-55$   & $-8\, C({\cal R}_{1/2})$& $-2$     \\
spin one              &  4 & 10    & $-22$  & $-170$  & $20 \, C(A)$            & $-8$     \\
a/s 2-form            &  6 & 30    & 132    & 420     & $-$                     & 23      \\
gravitino (W)         &$-12$& $-42$ & $-219$ & 1419    & $-$ & 50    \\
graviton              &  9 & 45    & 378    & $-2970$ & $-$                     & $-108$  \\
\hline
\end{tabular}}
\caption{\small 6D Results for Massless Fields, computed using
$R_{\scriptscriptstyle MN} = \frac{1}{2} \Lambda
g_{\scriptscriptstyle MN}$. The last column gives the result if
the spacetime is also maximally symmetric in 6 dimensions:
$R_{\scriptscriptstyle MNPQ} = (\Lambda/10) (g_{\scriptscriptstyle
MP} g_{\scriptscriptstyle NQ} - g_{\scriptscriptstyle NP}
g_{\scriptscriptstyle MQ})$.} \label{table 6D_Massless1}
\end{table}

Specializing to backgrounds with vanishing gauge fields and
constant dilaton field leads to vacuum space-times for which
$R_{MN} = 0$. It is then simple to see that the contributions to
the Gilkey coefficients of the gauge supermultiplet vanishes, with
the coefficients of the $\altRiem2$ and $\altf2$ terms both
cancelling between the gauge bosons and the gauginos. For the
gravity supermultiplet in these theories we also trivially have
$\Tr(a_0) = \Tr(a_1) = 0$ and
\beq
    \Tr_{I,{\rm het}}(a_2) = \frac{1}{180} \Bigl[155 - 191 + 28 + 7 + 1
    \Bigr] \altRiem2
    = 0 \,,
\eeq
again in agreement with ref.~\cite{Tseytlin}.

\subsubsection{Massive 10D Fields}

Massive 10D fields can arise in two ways in string theory. They
can arise as KK modes in the dimensional reduction of 11D
supergravity on a circle or a line segment, or as massive string
modes within the usual 10D string theories. Indeed, these two ways
are famously believed to be equivalent \cite{HoravaWitten}. The
contributions to the heat-kernel coefficients from various massive
10D fields are listed in Table \pref{10DtableM}.

\begin{table}
\centerline{
\begin{tabular}{ccc}
\hline
Multiplet & Particle Content & Number of States \\
\hline\hline
Hyper     &  2 spin 0 + 1 (symp-W) spin 1/2 & $2_B + 2_F$ \\
Gauge     &  1 spin 1 + 2 (symp-W) spin 1/2 & $4_B + 4_F$ \\
Tensor    &  1 spin 0 + 2 (symp-W) spin 1/2 +
1 (anti) self-dual 2-form & $4_B + 4_F$ \\
Gravitino &  1 (symp-W) spin 1/2 + 2 spin 1 + 1 (symp-W) spin 3/2 & $8_B + 8_F$ \\
Graviton   & 1 self-dual 2-form + 2 (symp-W) spin 3/2 + 1 spin 2 & $12_B + 12_F$ \\
\hline
\end{tabular}}
\caption{\small Particle content of massless 6D supermultiplets.}
\label{6Dmult-massless}

\vskip 0.5cm

\centerline{
\begin{tabular}{|l||c|c|}
\hline
& \multicolumn{2}{c|}{\Tr($a_2$)}\\
\cline{2-3}& $\frac{1}{48}\altRiem2$ & $\frac{1}{8}g_a^2 \altf2$ \\
\hline\hline
Hyper     & 1     & $-4C({\cal R}_h)$       \\
Gauge     & $-2$  & $8C(A)$                 \\
Tensor    & 10    & $-$                     \\
Gravitino & $-20$ & $16 C({\cal R}_{3/2})$  \\
Gravity   & 30    & $-$                     \\
\hline
\end{tabular}}
\caption{\small 6D results for massless supermultiplets, assuming
$\Lambda=0$. It is assumed that the tensor and graviton multiplets
do not carry the charge to which the background gauge fields
couple.} \label{table 6D_multiplets1}
\end{table}

A simple example which uses these results is the contribution of a
massive KK level which arises when the 11D theory is compactified
down to 10D on a circle. Writing the 10D indices as $\mu =
0,...,9$ and the 11th index as $s$, the 10D field content obtained
by dimensionally reducing in this case consists of the metric
components ($g_{\mu\nu}$, $g_{\mu s}$ and $g_{ss}$); the gravitino
components ($\psi_\mu$ and $\psi_s$); and the 3-form components
($C_{\mu\nu\lambda}$ and $C_{\mu\nu s}$). From the results of the
previous sections we see that these have the same field content as
a single massive 10D spin-2 particle, a single massive 10D
spin-3/2 particle and a single massive 3-form potential, and so
\beq
    \Tr_{10D-KK}(a_2) = \frac{1}{180} \Bigl[149 - 368 + 219
    \Bigr] \altRiem2
    = 0 \,.
\eeq

\begin{table}
\centerline{
\begin{tabular}{|l||c||c||c|c|c||c|}
\hline&\multicolumn{1}{c||}{$(-)^F$\tr($a_0$)}&\multicolumn{1}{c||}{$(-)^F$\tr($a_1$)}&\multicolumn{3}{c||}{$(-)^F$\tr($a_2$)}&\multicolumn{1}{c|}{$(-)^F$\tr($a_2$)$|_{ms}$ }\\
\cline{2-7}& 1 & $\frac{1}{10}\Lambda$ & $\frac{1}{360}\altRiem2$ & $\frac{1}{600}\Lambda^2$ & $\frac{1}{12}g_a^2 \altf2$ & $\frac{1}{25}\Lambda^2$ \\
\hline\hline
spin zero ($\xi=0$)   &  1  & $-5$  &  2     & 70      & $-C({\cal R}_0)$       &  3     \\
spin zero ($\xi=-1/5$)&  1  & 1     &  2     & $-2$    & $-C({\cal R}_0)$       &  0     \\
spin one-half (symp)  & $-4$& $-10$ &  7     & $-55$   & $-8C({\cal R}_{1/2})$  & $-2$   \\
spin one              &  5  & 5     & $-20$  & $-100$  & $19C(A)$               & $-5$   \\
a/s 2-form            & 10  & 40    &  110   & 250     & $-$                    & 15     \\
gravitino (symp)      &$-16$& $-80$ & $-212$ & $-220$  & $-32C({\cal R}_{3/2})$ & $-18$  \\
graviton              &  14 & 0     & 358    & $-1870$ & $-$                    & $-63$  \\
\hline
\end{tabular}}
\caption{\small 6D Results for Massive Fields, with $R_{\scriptscriptstyle MN} =
\frac{1}{2} \Lambda g_{\scriptscriptstyle MN}$. The last column gives the result if
the 6 dimensions are maximally symmetric.} \label{table
6D_Massive1}
\end{table}

\subsection{6D Examples}

In 6 dimensions there is a larger variety of supergravity theories
possible than in 10 dimensions, and so in this case we present our
results in terms of the various supermultiplets which are
encountered rather than attempting to independently list the most
commonly-occurring of the supergravities which are possible. Since
the particle content of a supermultiplet depends on whether or not
the particles are massless or massive, we treat each separately.
Since we allow $\Lambda \ne 0$ in Tables \pref{table 6D_Massless1}
and \pref{table 6D_Massive1}, in tabulating the Gilkey
coefficients for the gravitino we use the results from the
appendix.

\subsubsection{Massless Multiplets}

\noindent The contributions to the Gilkey coefficients which
result for massless particles in 6 dimensions are listed in Table
\pref{table 6D_Massless1}, and the field content of the commonly
occurring massless supermultiplets for 6D supersymmetry are listed
in Table \pref{6Dmult-massless}. For the case $\Lambda=0$, the
resulting nonzero heat-kernel coefficients for these multiplets
are given in Table \pref{table 6D_multiplets1}. In this table we
imagine that all of the particles in a given supermultiplet share
the same charge for the background gauge fields, which is true if
the relevant gauge symmetries commute with supersymmetry. Because
of this choice we also take the 2-form, gravitino and gauge fields
to be neutral under the gauge symmetry.

\begin{table}
\begin{center}
\begin{tabular}{|c|c|c|}
\hline& Multiplet & Field Equivalent \\
\hline\hline
Gauge & $\sixteenb_m$ &  ($A_M^m$,$2\psi^m$,3$\phi^m$)  \\
Gravitino & $\sixtyfourb_m$ &
  ($\psi_M^m$,2$A_{MN}^m$,2$A_M^m$,4$\psi^m$,2$\phi^m$) \\
Gravity & $\eightyb_m$ &
  ($g_{MN}^m$,2$\psi_M^m$,$A_{MN}^m$,3$A_M^m$,2$\psi^m$,$\phi^m$) \\
\hline
\end{tabular}
\end{center}
\caption{Massive representations of $(2,0)$ supersymmetry in 6
dimensions, labelled by their dimension. Note that the fermions in
this table are not chiral and the 2-form potentials are not
self-dual or anti-self-dual. The superscript `$m$' indicates the
corresponding field describes a massive particle (rather than
massless).} \label{massive6DReps}

\vskip 0.5cm

\centerline{
\begin{tabular}{|l||c|c|}
\hline
&\multicolumn{2}{c|}{\Tr($a_2$)}\\
\cline{2-3}& $\altRiem2$ & $\altf2$\\
\hline\hline
Gauge     &  0 & 0   \\
Gravitino &  0 & $-$ \\
Gravity   &  0 & $-$ \\
\hline
\end{tabular}}
\caption{\small 6D results for massive supermultiplets, assuming
$\Lambda=0$. It is assumed that the tensor and graviton multiplets
do not carry the charge to which the background gauge fields
couple.} \label{6Dmultiplets2}
\end{table}

\subsubsection{Massive Multiplets}

\noindent For massive 6D particles, the contributions to the
Gilkey coefficients found from the previous section are listed in
Table \pref{table 6D_Massive1}. The field content of the
commonly-occurring massive supermultiplets for 6D supersymmetry
are also listed in Table \pref{massive6DReps}. The resulting
heat-kernel coefficients for these multiplets are then given in
Table \pref{6Dmultiplets2} for the case $\Lambda=0$. In this table
we imagine that all of the particles in a given supermultiplet
share the same charge for the background gauge fields, which is
true if the relevant gauge symmetries commute with supersymmetry.
Because of this choice we also take the 2-form, gravitino and
gauge fields to be neutral under the gauge symmetry. Notice, in
particular, how $\Tr (a_2)$ vanishes for these 6D massive
multiplets provided the backgrounds are Ricci-flat ($\Lambda =
0$), as reported in a companion paper \cite{Doug}.

\subsection{4D Examples}

\begin{table}
\centerline{
\begin{tabular}{|l||c||c||c|c|c||c|}
\hline&\multicolumn{1}{c||}{$(-)^F$\tr($a_0$)}&\multicolumn{1}{c||}{$(-)^F$\tr($a_1$)}&\multicolumn{3}{c||}{$(-)^F$\tr($a_2$)}&\multicolumn{1}{c|}{$(-)^F$\tr($a_2$)$|_{ms}$ }\\
\cline{2-7}& 1 & $\frac{1}{3}\Lambda$ & $\frac{1}{720}\altRiem2$ & $\frac{1}{45}\Lambda^2$ & $\frac{1}{12}g_a^2 \altf2$ & $\frac{1}{270}\Lambda^2$ \\
\hline\hline
spin zero ($\xi=0$)   &  1 & $-2$  & 4      & 9      & $-C({\cal R}_0)$     & 58       \\
spin zero ($\xi=-1/6$)&  1 & 0     & 4      & $-1$   & $-C({\cal R}_0)$      & $-2$    \\
spin one-half (M)     &$-2$& $-2$  & 7      & $-3$   & $-4C({\cal R}_{1/2})$ & $-11$   \\
spin one              &  2 & 8     & $-52$  & $-12$  & $22C(A)$              & $-124$  \\
a/s 2-form            &  1 & $-2$  & 364    & 9      & $-$                   & 418     \\
gravitino (M)         &$-2$& $-18$ & $-233$ & 137    & $-4C({\cal R}_{3/2})$ & 589     \\
graviton              &  2 & 32    & 848    & $-522$ & $-$                   & $-2284$ \\
\hline
\end{tabular}}
\caption{\small 4D Results for Massless Fields, with $R_{\scriptscriptstyle MN} =
\Lambda g_{\scriptscriptstyle MN}$. The last column specializes to the
maximally-symmetric case, which in 4D implies
$R_{\scriptscriptstyle MNPQ} = (\Lambda/3) (g_{\scriptscriptstyle
MP} g_{\scriptscriptstyle NQ} - g_{\scriptscriptstyle NP}
g_{\scriptscriptstyle MQ})$.} \label{table 4D_Massless1}
\end{table}

There are considerably more supergravity theories possible in 4
dimensions than 6, and so we again list results as a function of
the particle content of 4D supermultiplets. As for the 6D case
this requires a separate discussion of the massless and massive
cases. A summary of the results of previous sections, specialized
to Einstein geometries, $R_{MN} = \Lambda \, g_{MN}$ is given in
Table \pref{table 4D_Massless1}. Since we allow $\Lambda$ to be
nonzero, we use the results in the appendix to tabulate the Gilkey
coefficients for the gravitino.

\subsubsection{Massless Multiplets}

\noindent The field content of the usual massless supermultiplets
for 4D supersymmetry are listed in Table \pref{4Dmult-massless}.
The corresponding nonzero heat-kernel coefficients for these
multiplets are given in Table \pref{4Dtable-sugra} for the case
$\Lambda=0$. If there is a nonzero cosmological constant, then as
discussed at the beginning of this section, there can be
additional $\Lambda$-dependent terms. We imagine that all of the
particles in a given supermultiplet share the same charge for the
background gauge fields. As usual we also take the 2-form,
gravitino and skew-tensor fields to be neutral under the
background gauge symmetry.

Although the contributions of 4D multiplets are typically nonzero,
they often give zero once they are summed over the particle
content of a multiplet of extended supersymmetry. For example,
combining one gauge multiplet with 3 conformally-coupled ($\xi =
-\frac16$) matter multiplets in the adjoint representation (${\cal
R}_m = A$) gives the field content of $N=4$ super-Yang Mills
theories. Specializing to flat space ($\Lambda = 0$) and summing
the appropriate entries in Table \pref{4Dtable-sugra} then
reproduces the famous result $\Tr(a_0) = \Tr(a_1) = \Tr(a_2) = 0$
for this combination.

\begin{table}
\centerline{
\begin{tabular}{ccc}
\hline
Multiplet & Particle Content & Number of States \\
\hline\hline
Matter    &  2 spin 0 + 1 (W) spin 1/2 & $2_B + 2_F$ \\
Gauge     &  1 spin 1 + 1 (W) spin 1/2 & $2_B + 2_F$ \\
Gravitino &  1 spin 1 + 1 (W) spin 3/2 & $2_B + 2_F$ \\
Gravity   &  1 (W) spin 3/2 + 1 spin 2 & $2_B + 2_F$ \\
\hline
\end{tabular}}
\caption{\small Particle content for $N=1$ massless supermultiplets in 4D.}
\label{4Dmult-massless}

\vskip 0.5cm

\centerline{
\begin{tabular}{|l||c|c|}
\hline& \multicolumn{2}{c|}{\Tr($a_2$)}\\
\cline{2-3} & $\frac{1}{48}\altRiem2$ & $\frac{1}{2}g_a^2 \altf2$ \\
\hline\hline
Matter    & 1     & $-C({\cal R}_m)$ \\
Gauge     & $-3$  & $3C(A)$          \\
Gravitino & $-19$ & $-$              \\
Gravity   & 41    & $-$              \\
\hline
\end{tabular}}
\caption{\small Results for massless supermultiplets in
4D. For the case $\Lambda \ne 0$, there will be additional $\Lambda$-dependent terms which we do not write.} \label{4Dtable-sugra}
\end{table}

\begin{table}
\centerline{
\begin{tabular}{|l||c||c||c|c|c||c|}
\hline&\multicolumn{1}{c||}{$(-)^F$\tr($a_0$)}&\multicolumn{1}{c||}{$(-)^F$\tr($a_1$)}&\multicolumn{3}{c||}{$(-)^F$\tr($a_2$)}&\multicolumn{1}{c|}{$(-)^F$\tr($a_2$)$|_{ms}$ }\\
\cline{2-7}& 1 & $\frac{1}{3}\Lambda$ & $\frac{1}{720} \altRiem2$ & $\frac{1}{45}\Lambda^2$ & $\frac{1}{12}g_a^2 \altf2$ & $\frac{1}{270}\Lambda^2$ \\
\hline\hline
spin zero ($\xi=0$)   &  1   & $-2$  &  4     & 9      & $-C({\cal R}_0)$      & 58      \\
spin zero ($\xi=-1/6$)&  1   & 0     &  4     & $-1$   & $-C({\cal R}_0)$      & $-2$    \\
spin one-half (M)     & $-2$ & $-2$  &  7     & $-3$   & $-4C({\cal R}_{1/2})$ & $-11$   \\
spin one              &  3   & 6     & $-48$  & $-3$   & $21C(A)$              & $-66$   \\
a/s 2-form            &  3   & 6     &  312   & $-3$   & $-$                   & 294     \\
gravitino (M)         & $-4$ & $-16$ & $-226$ & 24     & $-8C({\cal R}_{3/2})$ & $-82$   \\
graviton              &  5   & 20    &  800   & $-435$ & $-$                   & $-1810$ \\
\hline
\end{tabular}}
\caption{\small 4D Results for massive fields, with
$R_{\scriptscriptstyle MN} = \Lambda g_{\scriptscriptstyle MN}$.
The last column specializes to maximally-symmetric 4D background
geometries.} \label{table 4D_Massive}
\end{table}

\subsubsection{Massive Multiplets}

\noindent Finally, the heat-kernel coefficients for massive 4D
fields are given in Table \pref{table 4D_Massive}. These results
may then be assembled into massive representations of 4D
supersymmetry, as listed in Table \pref{4Dmult-massive}. The
corresponding heat-kernel coefficients for these multiplets are
given in Table \pref{4Dtable-sugramassive}, assuming all members
of the multiplet share the same mass (i.e., assuming $\Lambda=0$).
Again, if $\Lambda \ne 0$, then there can be additional
$\Lambda$-dependent terms which we do not follow. We take all of
the particles in a given supermultiplet to share the same charge
for the background gauge fields. The 2-form, gravitino and
skew-tensor fields are taken to be neutral under the background
gauge symmetry.

\begin{table}
\centerline{
\begin{tabular}{ccc}
\hline
Multiplet & Particle Content & Number of States \\
\hline\hline
Matter    & 2 spin 0   + 1 (M) spin 1/2                  & $2_B + 2_F$ \\
Gauge     & 1 spin 0   + 2 (M) spin 1/2 + 1 spin 1       & $4_B + 4_F$ \\
Gravitino & 1 (M) spin 1/2 + 2 spin 1   + 1 (M) spin 3/2 & $6_B + 6_F$ \\
Gravity   & 1 spin 1   + 2 (M) Spin 3/2 + 1 spin 2       & $8_B + 8_F$ \\
\hline
\end{tabular}}
\caption{\small Particle content for massive $N=1$ supermultiplets in 4D.}
\label{4Dmult-massive}

\vskip 0.5cm

\centerline{
\begin{tabular}{|l||c|c|}
\hline&\multicolumn{2}{c|}{\Tr($a_2$)}\\
\cline{2-3} & $\frac{1}{48} \altRiem2$ & $\frac{1}{2}g_a^2 \altf2$\\
\hline\hline
Matter    &  1    & $-C({\cal R}_m)$  \\
Gauge     & $-2$  & $2 \,C(A)$  \\
Gravitino & $-21$ & $-$  \\
Gravity   & $20$  & $-$  \\
\hline
\end{tabular}}
\caption{\small Results for massive supermultiplets in 4D. For the
case $\Lambda \ne 0$, there will be additional $\Lambda$-dependent terms which we do not write.}
\label{4Dtable-sugramassive}
\end{table}

\section{Conclusions}

This paper accomplishes several aims regarding one-loop
contributions to the effective action for a wide class of field
theories in a variety of dimensions.

First, we set up the quadratic part of the action for spins 0
through 2 in arbitrary spacetime dimensions in a way which is
useful for calculations. In particular, we set up a covariant
gauge for each spin which removes all mixings between fields that
transform differently under local Lorentz transformations. For
massive particles we show how to disentangle the higher-spin
fields from their lower-spin would-be Goldstone counterparts.

We then use this formulation to compute the leading ultraviolet
sensitivity which arises within a loop of any such particle. We
are able to do so because the gauge choice described above allows
us to use standard results for the heat-kernel coefficients for a
broad class of background fields. Finally, we tabulate these
coefficients for some of the fields and dimensions (4,6,10 and 11)
of particular interest for applications.

We expect the generality of our expressions to be useful for a
variety of future applications.

\section*{Acknowledgements}
We thank Arkady Tseytlin for helpful discussions. This research
was supported in part by a McGill University graduate fellowship
as well as research funds from N.S.E.R.C. (Canada), McMaster
University and from the Killam Foundation.

\appendix

\section{Appendix: Gravitini With $\Lambda \ne 0$}

\noindent In this appendix we slightly generalize the treatment of
massless and massive spin-3/2 particles given in the main text to
include the possibility that the lagrangian density includes a
nonzero cosmological constant (or a nontrivial scalar potential
once the background scalar field equations are satisfied). As
discussed in \S3, the nonzero cosmological constant implies
particles in a supermultiplet need no longer be degenerate in
mass, and so we calculate here how this effect plays out for the
gravitino. For instance, this case arises in four dimensions,
where an anti-de Sitter (AdS) cosmological constant term in the
action is not precluded by supersymmetry itself. Even though the
application of most interest is to four dimensions, we carry the
spacetime dimension $n$ as a variable in this appendix in case
more general applications of the expressions derived here should
become of interest.

\nisubsubsection{Massless Gravitino}

\noindent In this case we take the spin-2 field to be described by the
Einstein-Hilbert action supplemented by the cosmological term,
which in our conventions is
\beq
    \e \L_{EH} = -\frac{1}{2\kappa^2} (R-2\Lambda) \,.
\eeq
Supersymmetry then requires the lagrangian density for the
spin-3/2 particle to be described by
\beq
    \e \L_{VS} = -\frac{1}{2} \left( \overline{\psi}_M \Gamma^{MNP}
    D_N \psi_P -m_{3/2} \overline{\psi}_M \Gamma^{MN} \psi_N \right)
    \,,
\eeq
where we shall see how the parameter $m_{3/2}$ is related by
supersymmetry to the cosmological constant. The presence of this
`mass' term does not mean that supersymmetry is broken; rather it is
required in order to ensure that the gravitino/graviton action
remains gauge invariant.

The combined gravitino-graviton lagrangian is invariant under the
linearized supersymmetry transformations
\begin{eqnarray}
    \delta e_M^A &=& -\frac{\kappa}{4} \, \overline{\psi}_M \Gamma^A
    \epsilon + {\rm c.c.} \nn \\
    \label{eqn: psi_gauge2}
    \delta \psi_M &=& \frac{1}{\kappa} \left( D_M + \frac{1}{(n-2)}
    m_{3/2}
    \Gamma_M \right) \epsilon \,,
\end{eqnarray}
provided $m_{3/2}$ is related to $\Lambda$ by
\beq
    \label{Lambda}
    \Lambda = \frac{2(n-1)}{(n-2)} \, m_{3/2}^2.
\eeq
Notice that for any $n>2$ this requires $\Lambda > 0$, which in
our conventions corresponds to having anti-de Sitter space as the
maximally-symmetric background solution. In 4D this reduces to the
standard result $\Lambda_4 = 3 m_{3/2}^2$ \cite{sugra}.

To put the spin-3/2 lagrangian into a form for which the general
expressions for the Gilkey coefficients apply, we now use the
gauge-averaging term
\beq \label{3/2gf2}
    \e L_{VS}^{\, gf} = -\frac{1}{2 \, \xi_{3/2}} \,
    (\overline{\Gamma \cdot \psi})
    (\nott{D}+\gamma)(\Gamma \cdot \psi) \,.
\eeq
After making the field redefinition $\psi_M \rightarrow \psi_M + A
\Gamma_M \Gamma \cdot \psi$, we find that the following choices
for $A$, $\xi$, and $\gamma$
\beq
    A  = \frac{1}{2-n} \,, \qquad
    \xi^{-1}_{3/2} = \frac{2-n}{4} \,, \qquad
    \gamma = \left( \frac{n }{2-n} \right) \, m_{3/2} \,,
\eeq
lead to the an expression for the vector-spinor lagrangian given by
\beq
    \e (\L_{VS} + \L_{VS}^{\, gf}) = -\frac{1}{2} \, \,
    \overline{\psi}_M ( \nott{D}+m_{3/2})
    \psi^M \,.
\eeq

Following the analogous procedure in the main text, we obtain the
result for the vector-spinor field in the presence of a
cosmological constant:
\begin{eqnarray}
    \tr_{VS}(a_0) &=& \frac{n}{2} \, \cN_{3/2} \nn \\
    \tr_{VS}(a_1) &=& n \cN_{3/2} \left( \frac{1}{24}R -
    \frac{1}{2} m_{3/2}^2 \right) \nn \\
    \tr_{VS}(a_2) &=& \frac{\cN_{3/2}}{360} \left[
    \left(30-\frac{7n}{8} \right) \Riem2 - n \Ricci2 +
    \frac{5n}{8} R^2 + \frac{3n}{2} \Box R \right] \nn \\
          & & + \frac{n \d g_a^2}{12} \, C(\cR_{3/2})
          F^{{a} \hspace{9pt}}_{MN}
          F_{a \hspace{9pt}}^{MN}
          + \frac{n}{24} \cN_{3/2} \left( -m_{3/2}^2 R
          + 6 \, m_{3/2}^4 \right)
\end{eqnarray}
with $m_{3/2}^2$ defined by eq. ($\ref{Lambda}$).

The ghost action may be read from the supersymmetry transformation
rules, from which we see that $\delta( \Gamma \cdot \psi ) =
\frac{1}{\kappa}[ \nott{D} + \frac{n}{n-2}m_{3/2} ]
\epsilon$, and so we find two bosonic, Faddeev-Popov spinor ghosts
with the lagrangian
\beq
    \e \L_{LVFPgh} = -\overline{\omega}^i \left( \nott{D} + \frac{n \,
    m_{3/2}}{n-2} \right)
    \omega_i. \label{eqn: FadeevGhost2}
\eeq
This has the same form as the spin-1/2 lagrangian, eq. ($\ref{eqn:
L_spin1/2}$), although with a $\Lambda$-dependent mass. In order
to use this we require the following spin-1/2 results for the
Gilkey-DeWitt coefficients quoted in the main text, generalized to
include the fermion mass, $m^2$, inside $X$:
\begin{eqnarray}
    \label{eqn: a_spin1/2a}
    \tr_{1/2}(a_0) &=& \frac{\cN_{1/2}}{2} \nn \\
    \tr_{1/2}(a_1) &=& \frac{\cN_{1/2}}{24} (R-12 m^2) \nn \\
    \tr_{1/2}(a_2) &=& \frac{\cN_{1/2}}{360} \left[ -\frac{7}{8}
    \Riem2 - \Ricci2 + \frac{5}{8}
    (R-12m^2)^2 + \frac{3}{2} \Box R \right]
    \nn \\
    &&  \qquad\qquad + \frac{\d g_a^2}{12} \, C(\cR_{{1}/{2}}) \f2.
\end{eqnarray}
The Faddeev-Popov ghost result for $\tr[a_k]$ is then obtained by
multiplying these expressions by $-2$, and specializing to the
`mass' $m = n \, m_{3/2}/(n-2)$.

The use of the operator $(\nott{D} + \gamma)$ in the gauge-fixing
lagrangian, eq.~\pref{3/2gf2}, leads to a bosonic,
Nielsen-Kallosh ghost. Rewriting $\gamma$ in terms of $m_{3/2}$,
we see that the Nielsen-Kallosh ghost has the lagrangian
\beq
    \e \L_{LVNKgh} = -\overline{\omega} \left( \nott{D} - \frac{n \,
    m_{3/2}}{n-2} \right) \omega.
\eeq
This ghost therefore contributes $-1$ times the spin-1/2 result to
$\tr[a_k]$, with $m = -n m_{3/2}/(n-2)$.

Adding the vector-spinor result together with its associated ghosts, we
obtain the following contribution to $\tr[a_k]$ by physical
spin-3/2 states in the presence of a cosmological constant:
\begin{eqnarray}
    \tr_{3/2}(a_0) &=& \frac{\cN_{3/2}}{2} (n-3) \nn \\
    \tr_{3/2}(a_1) &=& \frac{\cN_{3/2}}{24} \left( (n-3)R -
    \frac{6n(n^2-7n+4)}{(n-1)(n-2)} \Lambda \right) \nn  \\
    \tr_{3/2}(a_2) &=& \frac{\cN_{3/2}}{360} \left[
    \left( 30-\frac{7}{8}(n-3) \right)
    \Riem2 - (n-3) \Ricci2 \nn \right. \\
    && \qquad  + \frac{5}{8}(n-3)R^2
     + \frac{3}{2} (n-3) \Box R - \frac{15n(n^2-7n+4)\Lambda R}{2(n-1)(n-2)}  \nn \\
    & & \qquad \left. +\frac{45n(n^4-11n^3+24n^2-32n+16)\Lambda^2}{2(n-1)^2(n-2)^2}
     \right] \nn \\
    && \qquad + \frac{\d g_a^2}{12}(n-3) \, C(\cR_{3/2}) \, \f2.
\end{eqnarray}

\nisubsubsection{Massive Gravitino}

\noindent This section follows closely the procedure outlined in the massive
spin-3/2 section of the main text. Starting from  eq.~\pref{LmVS}, which was
our ansatz for a massive spin-3/2 lagrangian, we again find that this lagrangian
can be made invariant under the supersymmetry transformations
\be
    \delta \psi_M = \frac{1}{\kappa} D_M
    \epsilon + \mu \Gamma_M \epsilon \qquad \hbox{and} \qquad
    \delta \chi = f \epsilon \,.
\ee
In this case, however, $f$ is given by
\be f^2 = (n-1)(n-2)\mu^2-\frac{\Lambda}{2\kappa^2}, \ee
while all other equations in eq.~\pref{params} remain unchanged. The dependence
of $f$ on $\Lambda$ is required in order to cancel the variation of the $\Lambda$
term in the Einstein-Hilbert action.

Again, following the procedure of the main text, we add a gauge-fixing term,
eq.~\pref{mVSgf}, and perform a field redefinition, eq.~\pref{redefns}, in order to put
the lagrangian into the form
\be
  \e (\L_{mVS} + \L_{mVS}^{\, gf}) = - {\overline{\psi}'}_M
  ( \nott{D} + {m'}_{3/2})
  {\psi'}^{M} - \overline{\chi}' (\nott{D} + {m'}_{1/2}) \chi'.
\ee
The parameters in the gauge-fixing lagrangian and in the field redefinitions can be
written in terms of $M$ and $\hat{M}$, defined as
\be
    M=(n-2)\kappa \mu \qquad {\rm and} \qquad
    \hat{M}=\sqrt{M^2+ \frac{2\Lambda}{n-2}}.
\ee
With these definitions, we find
\bea
     &&A=\sqrt{1+\beta^{\hspace{1pt} 2}}, \qquad B =-\frac{\beta}{2} \sqrt{n-2},
     \qquad C=-\hf, \qquad D=0, \nn \\
     &&\alpha = -\hf \sqrt{(n-2)(1+\beta^{\hspace{1pt} 2})}, \qquad \beta =
     \left[ \hf \left(\frac{n}{n-2}\right)\frac{M}{\hat{M}} - \hf \right]^{1/2}, \nn \\
     &&\qquad {m'}_{1/2}=- \hspace{1pt} \gamma=\hat{M}, \qquad {m'}_{3/2}=M.
\eea
For the case $\Lambda=0$, these expressions reduce to those given in eq.~\pref{mVS conditions}.
There is a possible subtlety in the above solution, which comes about because of our simplifying
assumption to take all free parameters to be real. We see that for certain choices of $M$ and
$\Lambda$, it's possible that some of the parameters will be imaginary. However, in the situations
for which our results apply we expect that $M \gg |\Lambda|$, and so in these cases this problem
will not arise.

From the gauge-fixing condition, we see that there are two Faddeev-Popov ghosts, each with mass
$\hat{M}$, and one Nielsen-Kallosh ghost, with mass $-\hat{M}$. The one loop
effective action for the ghosts is thus given by
\bea
    i\Sigma_{1/2} &=& \frac{1}{4} \Tr \log \left(\hat{M}^2-\nott{D}^2\right)  \nn \\
    &=& \frac{1}{4} \Tr \log \left(M^2 +\frac{2\Lambda}{n-2}-\nott{D}^2\right).
\eea
As usual, we factor the $M^2$ dependence out of our definition of $X$, and so obtain
\be
    X =  -\frac{1}{4}R+\frac{i}{2}\Gamma^{AB} F_{AB}^a t_a + \frac{2\Lambda}{n-2}.
\ee
The contribution to the Gilkey coefficients coming from the three ghosts is thus obtained by
multiplying eq.~\pref{eqn: a_spin1/2a} by $-3$, with $m^2=2\Lambda/(n-2)$. Similarly,
the Goldstone fermion contribution is also given by eq.~\pref{eqn: a_spin1/2a}, again with
$m^2=2\Lambda/(n-2)$. The contribution from the vector spinor is unchanged from the massless
case considered in the main text, and so its Gilkey coefficients are given by
eq.~\pref{eqn: a vs}. Summing these results, we arrive at the expression for a massive
gravitino in a background spacetime having nonzero cosmological constant:
\begin{eqnarray}
    \tr_{m3/2}(a_0) &=& \frac{\cN_{3/2}}{2} (n-2) \nn \\
    \tr_{m3/2}(a_1) &=& \frac{\cN_{3/2}}{24} \left( (n-2)R + \frac{48\Lambda}{n-2} \right)    \nn \\
    \tr_{m3/2}(a_2) &=& \frac{\cN_{3/2}}{360} \left[
    \left(30-\frac{7}{8}(n-2) \right)
    \Riem2 - (n-2) \Ricci2 \nonumber \right. \\
    && \qquad \left. + \frac{5}{8}(n-2)R^2 + \frac{3}{2} (n-2) \Box R
     +\frac{60\Lambda R}{(n-2)} - \frac{720\Lambda^2}{(n-2)^2} \right] \nn \\
    && \qquad + \frac{g_a^2}{12}(n-2) \d \, C(R_{3/2}) \f2 \,.
\end{eqnarray}

\end{document}